\documentclass[%
 reprint,
superscriptaddress,
 amsmath,amssymb,
 aps,
prl,
]{revtex4-2}

\usepackage{graphicx}
\usepackage{dcolumn}
\usepackage{bm}
\usepackage{hyperref}
\usepackage[normalem]{ulem}
\usepackage[caption=false]{subfig}
\usepackage{graphicx}
\usepackage{pict2e}
\usepackage[dvipsnames]{xcolor}
\usepackage{color,soul}

\newcommand\br{{\bf r}}
\newcommand\bv{{\bf v}}

\begin{document}

\title{Kinetic Monte Carlo Algorithms for Active Matter systems}

\author{Juliane U. Klamser}
\email{juliane.klamser@espci.psl.eu}
\affiliation{Gulliver UMR CNRS 7083, ESPCI Paris, Université PSL, 75005 Paris, France}%

\author{Olivier Dauchot}%
\email{olivier.dauchot@espci.fr}
\affiliation{Gulliver UMR CNRS 7083, ESPCI Paris, Université PSL, 75005 Paris, France}%

\author{Julien Tailleur}
\email{julien.tailleur@u-paris.fr}
\affiliation{Laboratoire Mati\`ere et Syst\`emes Complexes (MSC), UMR 7057 CNRS, Universit\'e de Paris, 75205 Paris, France}
\date{\today}

\begin{abstract}
  We study kinetic Monte Carlo (KMC) descriptions of active particles.
  We show that, when they rely on purely persistent, active steps,
  their continuous-time limit is ill-defined, leading to the vanishing
  of trademark behaviors of active matter such as the motility-induced
  phase separation, ratchet effects, as well as to a diverging
  mechanical pressure. We then show how, under an appropriate scaling,
  mixing passive steps with active ones leads to a well-defined
  continuous-time limit that however differs from standard active
  dynamics. Finally, we propose new KMC algorithms whose 
  continuous-time limits lead to the dynamics of active 
  Ornstein-Uhlenbeck, active Brownian, and run-and-tumble 
  particles.
\end{abstract}

\maketitle

Monte Carlo (MC) methods are widely popular across
disciplines~\cite{LandauBinder,RobertCasella}. At equilibrium,
detailed balance is enforced and unphysical dynamics can be used while
preserving the steady-state Boltzmann distribution.  Unphysical tricks
can then be exploited to accelerate equilibration without altering
steady-state averages of one-time observables, a property which has
led to many breakthroughs in equilibrium statistical
physics~\cite{WolfClusterAlgo1989,EquilibSwapAlgoBerthier2019,2DMeltingECMCBernard2011}.
MC algorithms have also been used to study diverse nonequilibrium
phenomena like
coarsening~\cite{GlauberDyn1963,Hastings1970,PESKUN1981}, slow
relaxation in disordered
systems~\cite{Berthier_2007,BerthierBiroliGlassRev2011,GlassAlgorithms2017},
granular media
\cite{Mueller1998,brey1999direct,montanero2000computer,cardenas2018contact},
self-assembly~\cite{NonEqSelfAssembly2018}, gel electrophoresis of DNA
\cite{NonEqDNA1997}, or surface
properties~\cite{BREEMAN1996195}. However, the relevance of
discrete-time dynamics for nonequilibrium systems is
questionable~\cite{SanzMarenduzzo10,Jabbari-FaroujiTrizac12,MCBookOfNewmanBarkema}
since no detailed-balance symmetry enforces a steady-state
distribution that is independent from the MC dynamics. This question
is particularly relevant in the field of active matter, where MC
simulations have been used extensively to simulate the collective
dynamics of active
particles~\cite{peruani2011traffic,thompson2011lattice,soto2014self,LevisBerthierKMC2014,BerthierAKMC2014,Levis_2015,sepulveda2016coarsening,manacorda2017lattice,KKK_AKMC2018,whitelam2018phase,kourbane2018exact,KKK_AKMC2019,shi2020self,ro2021disorder}.

Active matter constitutes a class of biological and synthetic systems
that are driven out of equilibrium at the microscopic
scale~\cite{marchetti2013hydrodynamics,bechinger2016active,romanczuk2012active}.
In their simplest form, they comprise assemblies of particles that
dissipate energy to exert self-propelling forces, hence breaking the
fluctuation-dissipation relation that would otherwise drive the
dynamics of passive colloids towards Boltzmann equilibrium. Active
systems have attracted a lot of attention due to their rich
phenomenologies, ranging from collective
motion~\cite{vicsek2012collective,chate2020dry}, to phase separation in
the absence of cohesive forces~ \cite{cates2015motility}, to
spatiotemporal chaos at zero Reynolds
number~\cite{wensink2012meso,stenhammar2017role,wu2017transition}.

The study of active matter systems is, however, challenging because of
two important limitations. Theoretically, first, there is no generic
expression for the steady-state distribution of active systems and no
counterpart to the Boltzmann weight to guide our
intuition. Numerically, then, studying the large-scale properties of
active systems requires sampling sizes much larger than the particle
persistence length. Defined as the typical distance a particle travels
before it forgets its initial orientation, the persistence length
often has to be much larger than the particle size for active matter
to display its most exciting features. This makes the system sizes to
be simulated much larger than for passive
systems~\cite{gregoire2004onset,weber2013long,solon2013revisiting,solon2015phase,shi2020self,ro2021disorder}.

To address this problem, a natural strategy would be, following the
success of MC in equilibrium, to replace the continuous-time setting
in which active dynamics are naturally defined by MC dynamics in which
time has been coarse grained. Several attempts along these lines have
been introduced recently, in particular to study
motility-induced phase separation (MIPS)~\cite{LevisBerthierKMC2014,KKK_AKMC2018}, the two-dimensional
melting~\cite{KKK_AKMC2018,KKK_AKMC2019,MeltingABP2018}, and
high-density binary mixtures~\cite{BerthierAKMC2014}. All these
approaches however suffer from a major drawback: {unlike for
  equilibrium systems, nothing guarantees that these MC dynamics, even
  in the proper limit, correspond to \textit{bona fide}
  continuous-time active dynamics.}

In this Letter, we bridge this gap by providing a class of active
kinetic MC dynamics (AKMC) whose continuous-time limit---which we
construct explicitly---is shown to encompass the celebrated
run-and-tumble (RT)~\cite{schnitzer1993theory,berg2008coli}, active
Brownian (AB)~\cite{fily2012athermal,farage2015effective} and active
Ornstein-Uhlenbeck (AOU)~\cite{szamel2014self,martin2021statistical}
dynamics.  To do so, we start by analyzing the continuous-time limit
of AKMC algorithms that have attracted a lot of attention
recently~\cite{LevisBerthierKMC2014,BerthierAKMC2014,Levis_2015,KKK_AKMC2018,MeltingABP2018,KKK_AKMC2019}. We
first show numerically and analytically that algorithms relying
exclusively on correlated, `active' steps lead to an ill-defined
continuous-time limit. We then show how the introduction of a finite
fraction of uncorrelated `passive' steps, together with a rescaling of
the propulsion speed, leads to a well-defined continuous active
dynamics. Importantly, the latter describes a new class of active
particles that differ from AB, RT, and AOU particles,
notwithstanding~\cite{LevisBerthierKMC2014,BerthierAKMC2014,Levis_2015}. We
close the Letter by discussing how our AKMC can be modified to lead to
RT, AB, and AOU particles hence providing a generic toolbox to simulate
active dynamics using AKMCs.

\textit{Active kinetic Monte Carlo dynamics---}  We consider a system of
$N$ active particles endowed with the following dynamics, adapted 
from~\cite{LevisBerthierKMC2014,KKK_AKMC2018}. At every
time step $t_n=n dt$, $N$ particles are successively chosen at random
and their positions $\br_i$ and self-propulsion velocities $\bv_i$ are
updated, in this order. A particle at $\br$ moves to a new position
$\br+\bv dt$, with probability
\begin{equation}
  f( \br, \bv dt) = \min\left[ 1, \exp\left( - \beta \Delta U(\br \to \br+\bv dt) \right) \right]\,,
\label{Eq:MetroFilterMin}
\end{equation}
where $\beta$ is a control parameter and $\Delta U = U(\br+\bv dt) -
U(\br)$ is the total energy change. Equation~\eqref{Eq:MetroFilterMin}
is nothing but a standard equilibrium Metropolis filter, in the
context of which $\beta$ would be the inverse temperature, and the
breakdown of detailed balance comes from the dynamics of $\bv$. A new
velocity $\bv(t_{n+1})$ is sampled from a Gaussian distribution
centered at $\bv(t_n)$, of standard deviation $\delta v=\sqrt{2 D_v
  dt}$, which is then folded back using reflecting boundary conditions
at $|\bv|=v_0$. (See Fig. S1 in~\cite{supp} for an illustration of
this procedure.) Successive particle displacements are thus correlated, 
hence leading
to a persistent motion characterized, in two space dimensions, by a
persistence time $\tau = \frac{v_0^2}{c^2 D_v}$ where $c$ is a
constant that can be computed exactly (see SM~\cite{supp}).

\begin{figure}[t!]
\centering
\subfloat{\label{fig1a}{\includegraphics{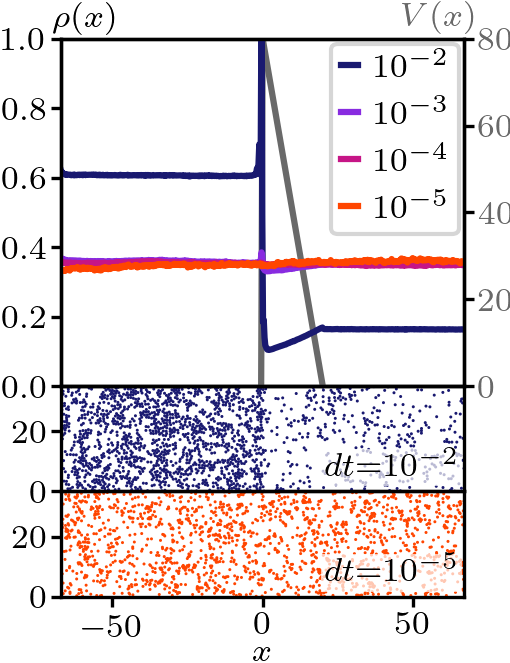}  }}%
\put(-120,0){(a) }
\subfloat{\label{fig1b}{\includegraphics{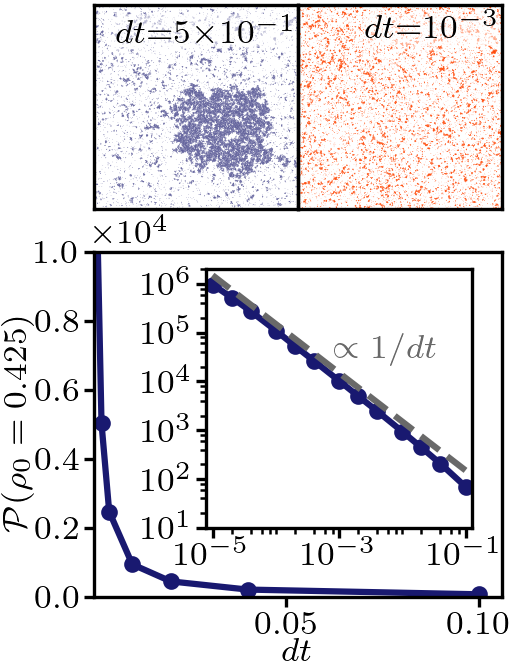}   }}%
\put(-125,110){(b)}
\put(-125,0){(c)}
\caption{Simulations of $N$ interacting active particles in $2D$ using
  the AKMC with $\beta = 1$. \textbf{(a)} As $dt$ is decreased from
  $10^{-2}$ to $10^{-5}$, the pumping effect induced by an asymmetric
  potential $V(x)$ (solid gray line, upper panel, explicit expression
  given in SM~\cite{supp}) disappears, as can be seen by comparing our
  simulations for $dt = 10^{-2}$ (snapshot in center panel and density
  profile shown as blue line in upper panel) and for $dt = 10^{-5}$
  (snapshot in lower panel and density profile shown as red line in
  upper panel).  Parameters: $\tau = 1$, $v_0 = 4$, average number
  density $\rho_0\simeq 0.36$; periodic (resp. closed) boundary
  conditions are used along $y$ (resp. $x$). \textbf{(b)} MIPS is
  observed for $dt = 0.5$ (blue, left panel) but not for $dt = 0.001$
  (red, right panel), using periodic boundary conditions.  Parameters:
  $\tau = 200$, $v_0 = 1$, $N = 43\,904$, $L=468$. \textbf{(c)} The
  mechanical pressure exerted on a confining potential
  $U_{\rm w}(x,y) = \frac{\Omega}{\nu}(x\pm x_{\rm
    w})^\nu\quad\text{for}\quad |x|>x_{\rm w}\;$ is measured using
  Eq.~\eqref{eq:defpressure} at bulk number density $\rho_0 = 0.425$
  using periodic boundaries along $y$.  It diverges as $1/dt$
  when $dt\to 0$. Parameters: $\tau = 10$, $v_0 = 1$, $L_y=32$,
  $x_{\rm w}=64$, $\Omega=10$, $\nu=8$.}\label{fig:alpha1}
\vspace{-5mm}  
\end{figure}

Figure~\ref{fig:alpha1} shows AKMC simulations of $N$ particles
interacting via a Weeks-Chandler-Andersen potential 
$U(r) = 4 \left[ (\sigma/r)^{12} -
  (\sigma/r)^{6}\right] + 1$ for $r< 2^{1/6}\sigma$ and $U(r)=0$
otherwise, with $\sigma = 2^{-1/6}$. Simulations are shown for
different time steps {$dt$}, keeping the self-propulsion speed $v_0$
and persistence time $\tau$ constant. Using large time steps, the
simulations reproduce standard features of active systems:
motility-induced phase separation~\cite{cates2015motility} is observed
and asymmetric obstacles are able to pump particles, hence generating
long-ranged perturbations to the density
field~\cite{galajda2007wall,wan2008rectification,tailleur2009sedimentation}.
However, both features disappear for smaller time steps.  Even more
surprising, the mechanical pressure exerted by the particles on a
confining potential $U_{\rm w}$, measured
as~\cite{GenActPresSolon2015}
\begin{equation}\label{eq:defpressure}
\mathcal{P}=\int_{x_{\rm bulk}}^{\infty} \rho(x) \partial_x U_{\rm w}(x)\,,
\end{equation}
is shown to diverge when $dt \to 0$. The AKMC algorithm introduced in
this section is thus not suitable to describe active dynamics.

\textit{Vanishing mobility in the continuous-time limit---} This
pathological behavior can be understood analytically by showing that
the particle mobility vanishes as $dt\to 0$, making the particles less
and less sensitive to forces other than the self-propulsion ones. Let
us consider the simpler problem of an isolated particle in the
presence of an external potential $U(x)$ in one space dimension.  The
generalization to higher dimensions and interacting particles is
straightforward. Reformulating the AKMC in one dimension leads to a
persistence time $\tau=4 v_0^2/(\pi^2 D_v) $~\cite{supp}. We denote
$P_n(x,v)$ the probability density to find the particle at position
$x$ with velocity $v$ at time $t_n$. Its evolution is given by
\begin{equation}
  P_{n+1}(x,v)=\int dx' dv' g(v|v') W(x|x', dt \, v')  P_{n}(x',v')
\end{equation}
where $g(v|v')$ is the probability density to transition from
self-propulsion velocity $v'$ to $v$ and where
\begin{eqnarray}\label{eq:W}
  W(x|x',\Delta x)&\equiv&f(x',\Delta x)\delta(x'+\Delta x -x) \notag\\
  &&+ [1-f(x',\Delta x)]\delta(x'-x)
\end{eqnarray}
is the probability density to transition from $x'$ to $x$.  The two
terms on the rhs of Eq.~\eqref{eq:W} correspond to hopping from
$x'\neq x$ into $x$ and to staying in $x'$, respectively.

The continuous-time limit of the evolution equation is obtained by
truncating the Kramers-Moyal
expansion of $\Delta
P\equiv P_{n+1}(x,v)- P_{n}(x,v)$ to first order in $dt$~\cite{gardiner1985handbook,van1992stochastic}. This has
been done with success for equilibrium MC dynamics---see
e.g.~\cite{KIKUCHI1991335,SanzMarenduzzo10,Jabbari-FaroujiTrizac12},
or~\cite{neuroevolution2020} for a nice application to neural
networks. As we show in the following, the generalization of this approach to the
active case leads to the Fokker-Planck
equation~\cite{Klamserfuture}:
\begin{equation}\label{eq:FPE}
\partial_t P(x,v;t)=- \frac{\partial}{\partial x} [v P(x,v;t)]+D_v \frac{\partial^2}{\partial v^2} P(x,v;t) \,,
\end{equation}
which is complemented by a zero-current condition 
\begin{equation}\label{eq:zero-current}
  \partial_v P(x, v; {t}) |_{v = \pm v_0} = 0\;.
\end{equation}
{The main lesson of this calculation} is that the confining potential drops out from
Eq.~\eqref{eq:FPE}.  To understand better how this happens, it is insightful to
write $\Delta P$ as
\begin{eqnarray}
  \Delta P&=& \int dv' g(v|v')[f(x-v' dt,v'dt) P_n(x-v'dt,v') \notag\\
  &&-f(x,v' dt)P_n(x,v')]\notag\\
  &&+\int dv' g(v|v') P_n(x,v') - P_n(x,v)\;.\label{eq:KM1}
\end{eqnarray}
Consider first the last line of Eq.~\eqref{eq:KM1}. Taylor expanding
$P_n(x,v')$ close to $v'=v$ leads to
\begin{equation*}
  \int dv' g(v|v') P_n(x,v') - P_n(x,v)=\sum_{k>0} \frac{a_k}{k!} \partial_v^kP_n(x,v)\,,
\end{equation*}
where $a_k$ is related to the $k^{\rm th}$ moment of the change in
velocity through $ a_k=(-1)^k \int dv' g(v|v') (v-v')^k$.  The
coefficient $a_1$ vanishes by symmetry in the $dt\to 0$ limit and
$a_2= \delta v^2$ provides the dominant order to $\Delta P$. This
confirms the scaling $\delta v= \sqrt{2 dt D_v}$ chosen above and
leads to the Laplacian on $v$ in Eq.~\eqref{eq:FPE}. The
zero-flux condition on $v$ is simply inherited from that of the
discrete-time process~\cite{gardiner1985handbook}. Consider now the
first two lines of Eq.~\eqref{eq:KM1}. To leading order in $dt$, they
are equivalent to
$- dt \int dv' g(v|v') v' \partial_x[f(x,v'dt) P_n(x,v')]$. This is
already of order $dt$ so that only the ${\cal O}(1)$ contribution of
the integral survives. To estimate the latter, we first note that
$\Delta U\simeq v dt U'(x)$ so that the Metropolis filter can be
approximated as
$f(x,v dt)\simeq 1-\beta v dt U'(x)\frac{1+{\rm sgn}(\Delta U)}2$. To
leading order, $f=1$ and the AKMC is insensitive to the filter in the
continuous-time limit. The computation can then be concluded by using
that $v'=(v'-v)+v$ and Taylor expanding $P(x,v')$ at $v'=v$, yielding
a leading order contribution $-dt v \partial_x
P(x,v)$. Mathematically, $U$ thus only enters Eq.~\eqref{eq:FPE}
at the next order in $dt$: the mobility of this AKMC vanishes linearly
in $dt$. Physically, $U$ is ignored by the particles since a succession
of infinitely small persistent steps lead to their systematic acceptance. 

The derivation above explains both the uniform distribution measured
in Fig.~\ref{fig:alpha1}a and the suppression of MIPS in
Fig.~\ref{fig:alpha1}b. Furthermore, as $dt \to 0$, particles
penetrate more and more into confining walls, so that the mechanical 
pressure exerted on the
walls, measured as Eq.~\eqref{eq:defpressure},
diverges.

\begin{figure}[t]
\centering
    \subfloat{\label{fig2a}{\includegraphics{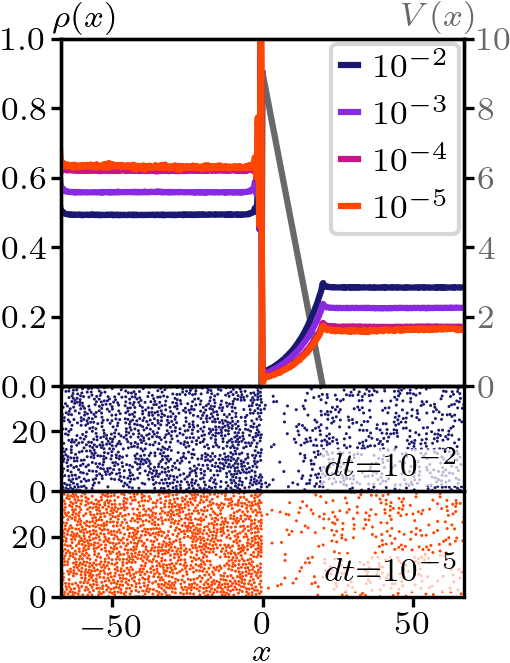}  }}%
    \put(-120,0){(a)}
    \subfloat{\label{fig2b}{\includegraphics{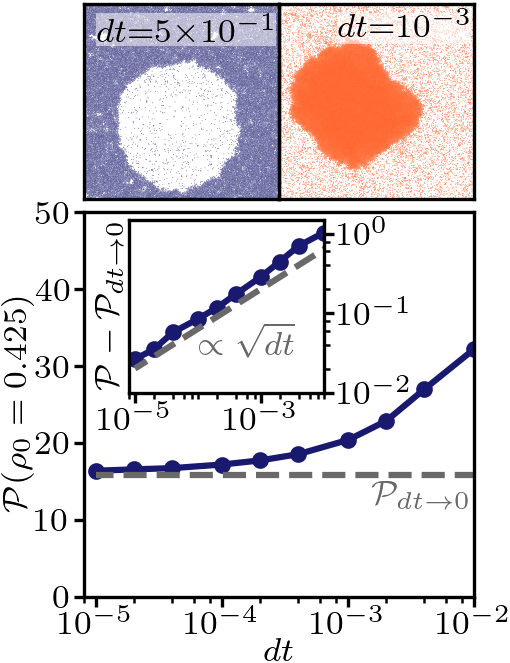}   }}%
    \put(-120,115){(b)}
    \put(-120,0){(c)}
    \caption{Simulations of $N$ interacting active particles using the
      blended AKMC with $\beta = 1$ { and $\alpha \in (0,1)$}.
      \textbf{(a)} As $dt$ is decreased from $10^{-2}$ to $10^{-5}$,
      the pumping effect induced by an asymmetric potential $V(x)$
      (gray line, upper panel, explicit expression given in
      SM~\cite{supp}) is now {converging to a stable nonequilibrium
        steady state}. Parameters: $\alpha = 0.4$, $\tau = 1$, $v_0 = 4$,
        $D = 1$, $\rho_0\simeq 0.36$. Periodic (resp. closed)
      boundary conditions are implemented along $y$
      (resp. $x$). \textbf{(b)} MIPS is now observed both for $dt = 0.5$
      (blue, left panel) and $dt = 0.001$ (red, right panel), using
      periodic boundary conditions. Parameters: $\alpha = 0.6$,
        $\tau = 200$, $v_0 = 1$, $D = 0.05$, $N = 43\,904$, $L=270$.
      \textbf{(c)} The mechanical pressure exerted on the
      confining potential $U_{\rm w}$ is measured using
        Eq.~\eqref{eq:defpressure} and has now a well-defined limit
      as $dt\to0$. Parameters: $\alpha = 0.6$, $\tau = 10$, $v_0 = 1$,
        $D = 0.05$.  The same geometry, wall potential, and
      densities are used as in
      Fig.~\ref{fig:alpha1}(c).}\label{fig:alpha06}
\vspace{-5mm}     
\end{figure}

\textit{A blended AKMC---} Since KMC algorithms admit a well-defined
continuous-time limit in
equilibrium~\cite{KIKUCHI1991335,SanzMarenduzzo10,Jabbari-FaroujiTrizac12,MCBookOfNewmanBarkema,neuroevolution2020},
it is natural to try and interpolate between passive and active KMC
dynamics~\cite{LevisBerthierKMC2014}.  To do so, we introduce a
blended AKMC as follows. At every time step, an attempt to move from
$x$ to $x+v dt/\alpha$ is done with probability $\alpha$ whereas a
move from $x$ to $x+\xi$ is attempted with probability $1-\alpha$,
where $\xi$ is sampled uniformly and independently at each time step
in $[-\sqrt{6 D dt/(1-\alpha)},\sqrt{6 D dt/(1-\alpha)}]$. In both
cases, the move is accepted or rejected using the
Metropolis filter defined in~\eqref{Eq:MetroFilterMin}. 
Note that the rescaling of the
propulsion speed and of the passive diffusivities with $\alpha$ 
will be proved below
to be crucial to the existence of an $\alpha$-independent well-defined
continuous-time limit. Figure~\ref{fig:alpha06} shows simulation
results for {$\alpha=0.4$ and $\alpha=0.6$}. Motility-induced phase
separation and a long-range modulation of the density field by an
asymmetric obstacle are again observed for large $dt$. This time,
however, these phenomena are stable as $dt\to 0$.  The mechanical
pressure exerted on confining walls also admits a well-defined limit.

The continuous-time limit of the blended AKMC can be constructed
analytically from the following extension of our calculation.  The
master equation now writes
\begin{eqnarray}\notag
  &&P_{n+1}(x,v)= \alpha \int dx' dv' g(v|v') W\Big(x\Big|x',{\frac{v' dt}{\alpha}}\Big)  P_{n}(x',v')\\
  &&+(1-\alpha)  \int dx' dv' d\xi g(v|v') W(x|x',\xi)  P_{n}(x',v') G(\xi)\,,\label{eq:MEblended}
\end{eqnarray}
where $G(\xi)$ is the uniform measure over $[-\sqrt{6 D
    dt/(1-\alpha)},\sqrt{6 D dt/(1-\alpha)}]$. By linearity, the
continuous-time limit of this blended AKMC is now readily
obtained. The first line of Eq.~\eqref{eq:MEblended} again leads to
the drift and diffusion terms derived in Eq.~\eqref{eq:FPE}{, albeit
  the latter multiplied by $\alpha$}.  The second line still leads to
the diffusion of the self-propulsion velocity multiplied by
$(1-\alpha)$, but also to the standard terms entering the
Fokker-Planck equation of a passive particle. All in all, this leads
to the Fokker-Planck equation
\begin{eqnarray}
\partial_t P_t(x, v)&& = - \partial_x \left[\left\{ v + \mu F(x)\right\} P_{t}(x, v)\right] +\nonumber\\
&&D\partial_x^2 P_t(x,v) + D_v \partial_v^2 P_{t}(x, v)
\label{Eq:FPEblended}
\end{eqnarray}
where Eq.~\eqref{Eq:FPEblended} is again supplemented by the
zero-current condition in Eq.~\eqref{eq:zero-current}.  This time, the
confining force $F(x) = -\partial_x U(x)$ survives in the $dt\to 0$
limit thanks to a finite mobility $\mu = \beta D$. Interestingly, comparing 
Eqs.~\eqref{eq:FPE} and \eqref{Eq:FPEblended} shows that the role 
played by the passive steps to restore the continuous-time limit
is not so much the introduction of translational diffusion as the
restoration of a finite mobility.

\begin{figure}[t]
\centering
\includegraphics{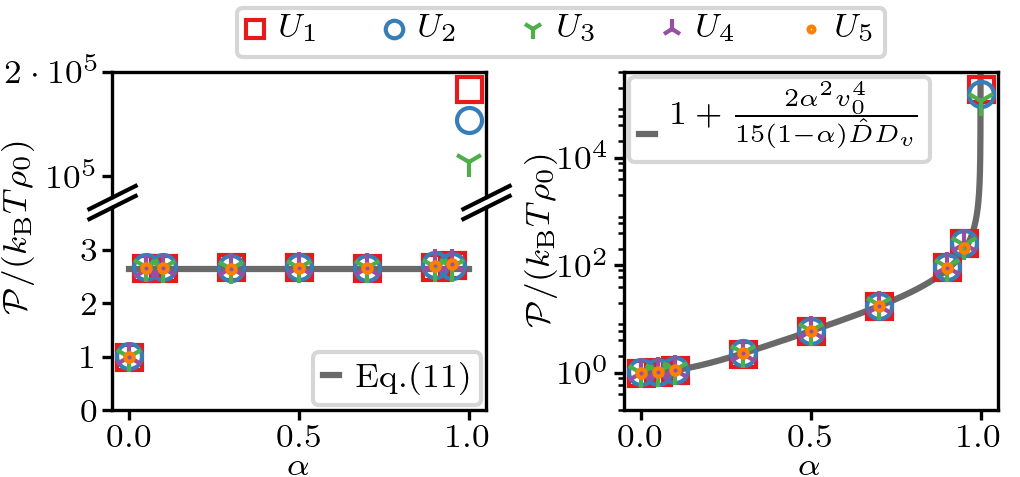}
\put(-240,0){(a)}
\put(-120,0){(b)}
\caption{\label{fig3} Mechanical pressure $\mathcal{P}$, defined in
  Eq.~\eqref{eq:defpressure} and normalized by its equilibrium value
  $\mathcal{P}=\rho_0 k_{\rm B}T$, measured as a function of the
  fraction of active steps $\alpha$ for noninteracting particles in
  1d, with $\beta = 1$, $v_0 = 1$, $\tau = 5$, $dt=10^{-4}$. 
  Symbols are measurements for several confining potentials
  $U_1$ to $U_5$, corresponding to the potential
  $U_{\rm w}$ defined in Fig.~\ref{fig:alpha1} with $(\nu,\Omega)$
  given by {$\{(8,10),\,(6,10),\,(4,10),\,(2,100),\,(2,10)\}$},
  respectively. The confining walls are located at
  $\pm x_\text{w}$.
  {\bf (a)} Simulations carried out with the blended AKMC for
  $\alpha\in(0,1)$, $D=1$, and $x_\text{w} = 15$. 
  The limiting cases $\alpha=0$ and $\alpha=1$
  correspond to purely passive and purely active KMCs, respectively.
  The solid {gray} line is the prediction of
  Eq.~\eqref{Eq:ContiTimeIdealPressure}. \textbf{(b)} Simulations
  carried out without rescaling the active steps and the passive
  diffusivity, using instead $x(t_{n+1})\to x(t_n) + v dt$ and
  $\xi\in[-\sqrt{6 \hat D dt},\sqrt{6 \hat D dt}]$, where $\hat{D}=1/6$ and $x_\text{w} = 5$.
  The lack of rescaling leads to an unphysical dependency of the 
  pressure on $\alpha$.}  
\end{figure}

We now compute the mechanical pressure predicted by
Eq.~\eqref{Eq:FPEblended} to check that the latter quantitatively
describes the small $dt$ limit of the blended AKMC. Integrating over
$v$ and using the zero-flux condition along $x$ imposed by the
confining wall leads to $\rho(x) U_{\rm w}'(x)=-\tfrac{D}{\mu}\rho'(x) + {\tfrac
{1} {\mu}} \bar v_1(x)$, where we define $\bar v_k(x)=\int dv P(x,v)
v^k$ and $\rho(x)=\bar v_0(x)$. Further integrating from $x=0$ to
$x=\infty$ leads to ${\mathcal P}=\frac D \mu \rho_0 + {\tfrac
{1} {\mu}} \int_0^\infty \bar v_1(x) dx$. To compute the last integral, we
multiply Eq.~\eqref{Eq:FPEblended} by $v^k$ and integrate over $v$ to
get, in the steady state,
\begin{equation}\label{vk}
   (k-1) \bar v_{k-2}=v_0^{k-1}[P(x,v_0)+(-1)^k P(x,-v_0)] +\partial_x \frac{J_k}{k  D_v}\;,
\end{equation}
where $J_k\equiv \bar v_{k+1} -\bar v_k \mu U_{\rm w}'(x)-D\partial_x
\bar v_k$. For $k=1$, Eq.~\eqref{vk} leads to $[P(x,v_0)-
  P(x,-v_0)]=-\partial_x J_1/D_v$. Injecting this into Eq.~\eqref{vk}
for $k=3$ and integrating both sides of the equation from $x=0$ to
$\infty$ leads to $6 D_v \int_0^\infty \bar v_1 dx = [3 v_0^2 \bar
  v_2(0)-\bar v_4(0)]$. Since the bulk of the system is homogeneous
and isotropic, $P(x=0,v)=\rho_0/(2v_0)$ and the mechanical pressure
reads
\begin{equation}
  {\mathcal{P}}  = \rho_0 \left(k_{\rm B}T +   \frac{2  v_0^4}{15 \mu D_v}\right)\;,
  \label{Eq:ContiTimeIdealPressure}
\end{equation}
where we have introduced $k_{\rm B}T\equiv \beta^{-1}$.
Figure~\ref{fig3}a shows the perfect match between
Eq.~\eqref{Eq:ContiTimeIdealPressure} and the mechanical
  pressure measured in numerical simulations for five different
potential stiffnesses and several values of {$\alpha\in(0,1)$}. For
  $\alpha < 1$, the pressure does not depend on the potential, which
indicates that the blended AKMC satisfies an equation of state in the
continuous-time limit. Note that the dependencies on $\alpha$ of the
active steps, ${\bf r}\to {\bf r}+{\bf v}dt/\alpha$, and of the
amplitude of the passive ones,
$\xi\in [-\sqrt{6D dt/(1-\alpha)},\sqrt{6D dt/(1-\alpha)}]$, may
look surprising at first glance---they were indeed absent in previous
AKMCs~\cite{LevisBerthierKMC2014,KKK_AKMC2018}. They are, however,
crucial to lead to continuous-time limits independent of $\alpha$,
as shown from Eq.~\eqref{Eq:FPEblended} and illustrated in
Fig.~\ref{fig3}.

\textit{AB, RT, and AOU algorithms.}  We have shown that our blended
AKMC leads to the Fokker-Planck equation~\eqref{Eq:FPEblended} in the
continuous-time limit. In two space dimensions, this active dynamics
is equivalent to the Langevin equation
\begin{equation}\label{eq:AKMCLangevin}
  \dot {\br}=  \bv -\mu \nabla U({\br}) + \sqrt{2 D}\boldsymbol\eta;\quad \dot\bv = \sqrt{2D_v} \boldsymbol \zeta
\end{equation}
where $\boldsymbol \eta$ and $\boldsymbol \zeta$ are two uncorrelated
unitary Gaussian white noises and $\bv$ experiences reflecting boundary
conditions at $|\bv|=v_0$. Interestingly, the dynamics of $\bv$
corresponds to none of the standard active particle models. As we now
show, our blended AKMC can be adapted to yield discrete-time versions
of AB, RT and AOU particles by solely modifying the dynamics of the
self-propulsion speed.  For RT and AB dynamics, the self-propulsion
speed ${\bf v}(t_n)$ lives on a circle of radius $v_0$ and is
parametrized by an angle $\theta(t_n)$. A discretized RT dynamics with
tumbling rate $\gamma$ is obtained by choosing
$\theta(t_{n+1})=\theta(t_{n})$ with probability $(1-\gamma dt)$ and
by sampling $\theta(t_{n+1})$ uniformly in $[0,2\pi)$ with probability
$\gamma dt$. To implement an AB dynamics with rotational diffusivity
$D_r$, $\theta(t_{n+1})$ is sampled from a wrapped Gaussian
distribution of standard deviation $\delta \theta=\sqrt{2 D_r dt}$,
centered at $\theta(t_{n})$. Finally, the AOU dynamics can be
implemented as follows.  A change of velocity ${\bf \delta v}(t_n)$ is
sampled uniformly in
$[-\sqrt{6 D_{\rm a} dt/(\tau^2)},\sqrt{6 D_{\rm a} dt/(\tau^2)}]^2$. It is
accepted with probability
\begin{equation}
  p=\text{min}\Big[1,\exp\Big(-\frac{\tau^2}{D_{\rm a}} \Delta U_v[{\bf v}(t_n)\to {\bf v}(t_n)+ {\bf \delta v}(t_n)]\Big)\Big]
\end{equation}
where $U_v({\bf v})=\frac 1 {2\tau} {\bf v}^2$. Carrying out the
continuous-time limit of the dynamics indeed leads to the
Fokker-Planck equation equivalent to
$\dot {\bf r}=  {\bf v}+\mu {\bf F} + \sqrt{2 D}
\boldsymbol{\eta}$ and
$\tau \dot{\bf v}=-{\bf v} + \sqrt{2 D_{\rm a}} \boldsymbol\zeta$ where
$\boldsymbol \eta$ and $\boldsymbol \zeta$ are two uncorrelated unitary
Gaussian white noises. 

Altogether we have shown how mixing passive steps with active ones
endow AKMCs with \textit{bona fide} continuous-time limits which
encompass the workhorse models of active matter. By clarifying the
connection between discrete and continuous-time dynamics, we believe
our work will trigger a wider use of AKMCs in active matter. They
should prove especially useful in the high density limit where
Langevin equations are particularly difficult to use. This regime has
indeed attracted a lot of attention
recently~\cite{henkes2011active,flenner2016nonequilibrium,berthier2019glassy,mandal2020extreme},
in particular due to its relevance to the modeling of confluent
tissues~\cite{matoz2017cell,loewe2020solid,henkes2020dense}, but also
because of the emergence of nontrivial spatial velocity
correlations~\cite{CapriniPRE20,CapriniSM21,szamel2021longranged}.
Finally, it would be interesting to generalize the approach developed
in this Letter to MC algorithms in which space has also been
discretized, which have recently attracted a lot of
attention~\cite{peruani2011traffic,thompson2011lattice,soto2014self,sepulveda2016coarsening,manacorda2017lattice,whitelam2018phase,kourbane2018exact,shi2020self}.

\acknowledgments{} JT acknowledges the financial support of ANR Grant
THEMA. The authors benefited from participation in the 2020 KITP
program on Active Matter supported by the Grant NSF PHY-1748958.

\bibliography{biblio}

\providecommand{\noopsort}[1]{}\providecommand{\singleletter}[1]{#1}%
\begin{thebibliography}{75}%
\makeatletter
\providecommand \@ifxundefined [1]{%
 \@ifx{#1\undefined}
}%
\providecommand \@ifnum [1]{%
 \ifnum #1\expandafter \@firstoftwo
 \else \expandafter \@secondoftwo
 \fi
}%
\providecommand \@ifx [1]{%
 \ifx #1\expandafter \@firstoftwo
 \else \expandafter \@secondoftwo
 \fi
}%
\providecommand \natexlab [1]{#1}%
\providecommand \enquote  [1]{``#1''}%
\providecommand \bibnamefont  [1]{#1}%
\providecommand \bibfnamefont [1]{#1}%
\providecommand \citenamefont [1]{#1}%
\providecommand \href@noop [0]{\@secondoftwo}%
\providecommand \href [0]{\begingroup \@sanitize@url \@href}%
\providecommand \@href[1]{\@@startlink{#1}\@@href}%
\providecommand \@@href[1]{\endgroup#1\@@endlink}%
\providecommand \@sanitize@url [0]{\catcode `\\12\catcode `\$12\catcode
  `\&12\catcode `\#12\catcode `\^12\catcode `\_12\catcode `\%12\relax}%
\providecommand \@@startlink[1]{}%
\providecommand \@@endlink[0]{}%
\providecommand \url  [0]{\begingroup\@sanitize@url \@url }%
\providecommand \@url [1]{\endgroup\@href {#1}{\urlprefix }}%
\providecommand \urlprefix  [0]{URL }%
\providecommand \Eprint [0]{\href }%
\providecommand \doibase [0]{https://doi.org/}%
\providecommand \selectlanguage [0]{\@gobble}%
\providecommand \bibinfo  [0]{\@secondoftwo}%
\providecommand \bibfield  [0]{\@secondoftwo}%
\providecommand \translation [1]{[#1]}%
\providecommand \BibitemOpen [0]{}%
\providecommand \bibitemStop [0]{}%
\providecommand \bibitemNoStop [0]{.\EOS\space}%
\providecommand \EOS [0]{\spacefactor3000\relax}%
\providecommand \BibitemShut  [1]{\csname bibitem#1\endcsname}%
\let\auto@bib@innerbib\@empty
\bibitem [{\citenamefont {Landau}\ and\ \citenamefont
  {Binder}(2014)}]{LandauBinder}%
  \BibitemOpen
  \bibfield  {author} {\bibinfo {author} {\bibfnamefont {D.}~\bibnamefont
  {Landau}}\ and\ \bibinfo {author} {\bibfnamefont {K.}~\bibnamefont
  {Binder}},\ }\href@noop {} {\emph {\bibinfo {title} {A Guide to Monte Carlo
  Simulations in Statistical Physics}}}\ (\bibinfo  {publisher} {Cambridge
  University Press},\ \bibinfo {year} {2014})\BibitemShut {NoStop}%
\bibitem [{\citenamefont {Robert}\ and\ \citenamefont
  {Casella}(2004)}]{RobertCasella}%
  \BibitemOpen
  \bibfield  {author} {\bibinfo {author} {\bibfnamefont {C.~P.}\ \bibnamefont
  {Robert}}\ and\ \bibinfo {author} {\bibfnamefont {G.}~\bibnamefont
  {Casella}},\ }\href@noop {} {\emph {\bibinfo {title} {Monte Carlo statistical
  methods.}}}\ (\bibinfo  {publisher} {Springer-Verlag, New York},\ \bibinfo
  {year} {2004})\BibitemShut {NoStop}%
\bibitem [{\citenamefont {Wolff}(1989)}]{WolfClusterAlgo1989}%
  \BibitemOpen
  \bibfield  {author} {\bibinfo {author} {\bibfnamefont {U.}~\bibnamefont
  {Wolff}},\ }\bibfield  {title} {\bibinfo {title} {Collective monte carlo
  updating for spin systems},\ }\href@noop {} {\bibfield  {journal} {\bibinfo
  {journal} {Phys. Rev. Lett.}\ }\textbf {\bibinfo {volume} {62}},\ \bibinfo
  {pages} {361} (\bibinfo {year} {1989})}\BibitemShut {NoStop}%
\bibitem [{\citenamefont {Berthier}\ \emph
  {et~al.}(2019{\natexlab{a}})\citenamefont {Berthier}, \citenamefont
  {Flenner}, \citenamefont {Fullerton}, \citenamefont {Scalliet},\ and\
  \citenamefont {Singh}}]{EquilibSwapAlgoBerthier2019}%
  \BibitemOpen
  \bibfield  {author} {\bibinfo {author} {\bibfnamefont {L.}~\bibnamefont
  {Berthier}}, \bibinfo {author} {\bibfnamefont {E.}~\bibnamefont {Flenner}},
  \bibinfo {author} {\bibfnamefont {C.~J.}\ \bibnamefont {Fullerton}}, \bibinfo
  {author} {\bibfnamefont {C.}~\bibnamefont {Scalliet}},\ and\ \bibinfo
  {author} {\bibfnamefont {M.}~\bibnamefont {Singh}},\ }\bibfield  {title}
  {\bibinfo {title} {Efficient swap algorithms for molecular dynamics
  simulations of equilibrium supercooled liquids},\ }\href@noop {} {\bibfield
  {journal} {\bibinfo  {journal} {J. Stat. Mech. Theor. Exp.}\ }\textbf
  {\bibinfo {volume} {2019}},\ \bibinfo {pages} {064004} (\bibinfo {year}
  {2019}{\natexlab{a}})}\BibitemShut {NoStop}%
\bibitem [{\citenamefont {Bernard}\ and\ \citenamefont
  {Krauth}(2011)}]{2DMeltingECMCBernard2011}%
  \BibitemOpen
  \bibfield  {author} {\bibinfo {author} {\bibfnamefont {E.~P.}\ \bibnamefont
  {Bernard}}\ and\ \bibinfo {author} {\bibfnamefont {W.}~\bibnamefont
  {Krauth}},\ }\bibfield  {title} {\bibinfo {title} {Two-step melting in two
  dimensions: First-order liquid-hexatic transition},\ }\href@noop {}
  {\bibfield  {journal} {\bibinfo  {journal} {Phys. Rev. Lett.}\ }\textbf
  {\bibinfo {volume} {107}},\ \bibinfo {pages} {155704} (\bibinfo {year}
  {2011})}\BibitemShut {NoStop}%
\bibitem [{\citenamefont {Glauber}(1963)}]{GlauberDyn1963}%
  \BibitemOpen
  \bibfield  {author} {\bibinfo {author} {\bibfnamefont {R.~J.}\ \bibnamefont
  {Glauber}},\ }\bibfield  {title} {\bibinfo {title} {Time‐dependent
  statistics of the ising model},\ }\href@noop {} {\bibfield  {journal}
  {\bibinfo  {journal} {J. Math. Phys.}\ }\textbf {\bibinfo {volume} {4}},\
  \bibinfo {pages} {294} (\bibinfo {year} {1963})}\BibitemShut {NoStop}%
\bibitem [{\citenamefont {Hastings}(1970)}]{Hastings1970}%
  \BibitemOpen
  \bibfield  {author} {\bibinfo {author} {\bibfnamefont {W.~K.}\ \bibnamefont
  {Hastings}},\ }\bibfield  {title} {\bibinfo {title} {Monte carlo sampling
  methods using markov chains and their applications},\ }\href@noop {}
  {\bibfield  {journal} {\bibinfo  {journal} {Biometrika}\ }\textbf {\bibinfo
  {volume} {57}},\ \bibinfo {pages} {97} (\bibinfo {year} {1970})}\BibitemShut
  {NoStop}%
\bibitem [{\citenamefont {Peskun}(1981)}]{PESKUN1981}%
  \BibitemOpen
  \bibfield  {author} {\bibinfo {author} {\bibfnamefont {P.}~\bibnamefont
  {Peskun}},\ }\bibfield  {title} {\bibinfo {title} {Guidelines for choosing
  the transition matrix in monte carlo methods using markov chains},\
  }\href@noop {} {\bibfield  {journal} {\bibinfo  {journal} {J. Comput. Phys.}\
  }\textbf {\bibinfo {volume} {40}},\ \bibinfo {pages} {327} (\bibinfo {year}
  {1981})}\BibitemShut {NoStop}%
\bibitem [{\citenamefont {Berthier}\ and\ \citenamefont
  {Kob}(2007)}]{Berthier_2007}%
  \BibitemOpen
  \bibfield  {author} {\bibinfo {author} {\bibfnamefont {L.}~\bibnamefont
  {Berthier}}\ and\ \bibinfo {author} {\bibfnamefont {W.}~\bibnamefont {Kob}},\
  }\bibfield  {title} {\bibinfo {title} {The monte carlo dynamics of a binary
  lennard-jones glass-forming mixture},\ }\href@noop {} {\bibfield  {journal}
  {\bibinfo  {journal} {J. Phys. Condens. Matter}\ }\textbf {\bibinfo {volume}
  {19}},\ \bibinfo {pages} {205130} (\bibinfo {year} {2007})}\BibitemShut
  {NoStop}%
\bibitem [{\citenamefont {Berthier}\ and\ \citenamefont
  {Biroli}(2011)}]{BerthierBiroliGlassRev2011}%
  \BibitemOpen
  \bibfield  {author} {\bibinfo {author} {\bibfnamefont {L.}~\bibnamefont
  {Berthier}}\ and\ \bibinfo {author} {\bibfnamefont {G.}~\bibnamefont
  {Biroli}},\ }\bibfield  {title} {\bibinfo {title} {Theoretical perspective on
  the glass transition and amorphous materials},\ }\href@noop {} {\bibfield
  {journal} {\bibinfo  {journal} {Rev. Mod. Phys.}\ }\textbf {\bibinfo {volume}
  {83}},\ \bibinfo {pages} {587} (\bibinfo {year} {2011})}\BibitemShut
  {NoStop}%
\bibitem [{\citenamefont {Ninarello}\ \emph {et~al.}(2017)\citenamefont
  {Ninarello}, \citenamefont {Berthier},\ and\ \citenamefont
  {Coslovich}}]{GlassAlgorithms2017}%
  \BibitemOpen
  \bibfield  {author} {\bibinfo {author} {\bibfnamefont {A.}~\bibnamefont
  {Ninarello}}, \bibinfo {author} {\bibfnamefont {L.}~\bibnamefont
  {Berthier}},\ and\ \bibinfo {author} {\bibfnamefont {D.}~\bibnamefont
  {Coslovich}},\ }\bibfield  {title} {\bibinfo {title} {Models and algorithms
  for the next generation of glass transition studies},\ }\href@noop {}
  {\bibfield  {journal} {\bibinfo  {journal} {Phys. Rev. X}\ }\textbf {\bibinfo
  {volume} {7}},\ \bibinfo {pages} {021039} (\bibinfo {year}
  {2017})}\BibitemShut {NoStop}%
\bibitem [{\citenamefont {M{\"u}ller}\ and\ \citenamefont
  {Herrmann}(1998)}]{Mueller1998}%
  \BibitemOpen
  \bibfield  {author} {\bibinfo {author} {\bibfnamefont {M.}~\bibnamefont
  {M{\"u}ller}}\ and\ \bibinfo {author} {\bibfnamefont {H.~J.}\ \bibnamefont
  {Herrmann}},\ }\bibinfo {title} {Dsmc --- a stochastic algorithm for granular
  matter},\ in\ \href {https://doi.org/10.1007/978-94-017-2653-5_30} {\emph
  {\bibinfo {booktitle} {Physics of Dry Granular Media}}},\ \bibinfo {editor}
  {edited by\ \bibinfo {editor} {\bibfnamefont {H.~J.}\ \bibnamefont
  {Herrmann}}, \bibinfo {editor} {\bibfnamefont {J.-P.}\ \bibnamefont {Hovi}},\
  and\ \bibinfo {editor} {\bibfnamefont {S.}~\bibnamefont {Luding}}}\ (\bibinfo
   {publisher} {Springer Netherlands},\ \bibinfo {address} {Dordrecht},\
  \bibinfo {year} {1998})\ pp.\ \bibinfo {pages} {413--420}\BibitemShut
  {NoStop}%
\bibitem [{\citenamefont {Brey}\ and\ \citenamefont
  {Ruiz-Montero}(1999)}]{brey1999direct}%
  \BibitemOpen
  \bibfield  {author} {\bibinfo {author} {\bibfnamefont {J.~J.}\ \bibnamefont
  {Brey}}\ and\ \bibinfo {author} {\bibfnamefont {M.}~\bibnamefont
  {Ruiz-Montero}},\ }\bibfield  {title} {\bibinfo {title} {Direct monte carlo
  simulation of dilute granular flow},\ }\href@noop {} {\bibfield  {journal}
  {\bibinfo  {journal} {Computer physics communications}\ }\textbf {\bibinfo
  {volume} {121}},\ \bibinfo {pages} {278} (\bibinfo {year}
  {1999})}\BibitemShut {NoStop}%
\bibitem [{\citenamefont {Montanero}\ and\ \citenamefont
  {Santos}(2000)}]{montanero2000computer}%
  \BibitemOpen
  \bibfield  {author} {\bibinfo {author} {\bibfnamefont {J.~M.}\ \bibnamefont
  {Montanero}}\ and\ \bibinfo {author} {\bibfnamefont {A.}~\bibnamefont
  {Santos}},\ }\bibfield  {title} {\bibinfo {title} {Computer simulation of
  uniformly heated granular fluids},\ }\href@noop {} {\bibfield  {journal}
  {\bibinfo  {journal} {Granular Matter}\ }\textbf {\bibinfo {volume} {2}},\
  \bibinfo {pages} {53} (\bibinfo {year} {2000})}\BibitemShut {NoStop}%
\bibitem [{\citenamefont {C{\'a}rdenas-Barrantes}\ \emph
  {et~al.}(2018)\citenamefont {C{\'a}rdenas-Barrantes}, \citenamefont
  {Mu{\~n}oz},\ and\ \citenamefont {Oquendo}}]{cardenas2018contact}%
  \BibitemOpen
  \bibfield  {author} {\bibinfo {author} {\bibfnamefont {M.~A.}\ \bibnamefont
  {C{\'a}rdenas-Barrantes}}, \bibinfo {author} {\bibfnamefont {J.~D.}\
  \bibnamefont {Mu{\~n}oz}},\ and\ \bibinfo {author} {\bibfnamefont {W.~F.}\
  \bibnamefont {Oquendo}},\ }\bibfield  {title} {\bibinfo {title} {Contact
  forces distribution for a granular material from a monte carlo study on a
  single grain},\ }\href@noop {} {\bibfield  {journal} {\bibinfo  {journal}
  {Granular Matter}\ }\textbf {\bibinfo {volume} {20}},\ \bibinfo {pages} {1}
  (\bibinfo {year} {2018})}\BibitemShut {NoStop}%
\bibitem [{\citenamefont {Bisker}\ and\ \citenamefont
  {England}(2018)}]{NonEqSelfAssembly2018}%
  \BibitemOpen
  \bibfield  {author} {\bibinfo {author} {\bibfnamefont {G.}~\bibnamefont
  {Bisker}}\ and\ \bibinfo {author} {\bibfnamefont {J.~L.}\ \bibnamefont
  {England}},\ }\bibfield  {title} {\bibinfo {title} {Nonequilibrium
  associative retrieval of multiple stored self-assembly targets},\ }\href@noop
  {} {\bibfield  {journal} {\bibinfo  {journal} {Proceedings of the National
  Academy of Sciences}\ }\textbf {\bibinfo {volume} {115}},\ \bibinfo {pages}
  {E10531} (\bibinfo {year} {2018})}\BibitemShut {NoStop}%
\bibitem [{\citenamefont {Barkema}\ and\ \citenamefont
  {Newman}(1997)}]{NonEqDNA1997}%
  \BibitemOpen
  \bibfield  {author} {\bibinfo {author} {\bibfnamefont {G.}~\bibnamefont
  {Barkema}}\ and\ \bibinfo {author} {\bibfnamefont {M.}~\bibnamefont
  {Newman}},\ }\bibfield  {title} {\bibinfo {title} {The repton model of gel
  electrophoresis},\ }\href@noop {} {\bibfield  {journal} {\bibinfo  {journal}
  {Physica A}\ }\textbf {\bibinfo {volume} {244}},\ \bibinfo {pages} {25}
  (\bibinfo {year} {1997})}\BibitemShut {NoStop}%
\bibitem [{\citenamefont {Breeman}\ \emph {et~al.}(1996)\citenamefont
  {Breeman}, \citenamefont {Barkema}, \citenamefont {Langelaar},\ and\
  \citenamefont {Boerma}}]{BREEMAN1996195}%
  \BibitemOpen
  \bibfield  {author} {\bibinfo {author} {\bibfnamefont {M.}~\bibnamefont
  {Breeman}}, \bibinfo {author} {\bibfnamefont {G.}~\bibnamefont {Barkema}},
  \bibinfo {author} {\bibfnamefont {M.}~\bibnamefont {Langelaar}},\ and\
  \bibinfo {author} {\bibfnamefont {D.}~\bibnamefont {Boerma}},\ }\bibfield
  {title} {\bibinfo {title} {Computer simulation of metal-on-metal epitaxy},\
  }\href@noop {} {\bibfield  {journal} {\bibinfo  {journal} {Thin Solid Films}\
  }\textbf {\bibinfo {volume} {272}},\ \bibinfo {pages} {195} (\bibinfo {year}
  {1996})}\BibitemShut {NoStop}%
\bibitem [{\citenamefont {Sanz}\ and\ \citenamefont
  {Marenduzzo}(2010)}]{SanzMarenduzzo10}%
  \BibitemOpen
  \bibfield  {author} {\bibinfo {author} {\bibfnamefont {E.}~\bibnamefont
  {Sanz}}\ and\ \bibinfo {author} {\bibfnamefont {D.}~\bibnamefont
  {Marenduzzo}},\ }\bibfield  {title} {\bibinfo {title} {Dynamic monte carlo
  versus brownian dynamics: A comparison for self-diffusion and crystallization
  in colloidal fluids},\ }\href@noop {} {\bibfield  {journal} {\bibinfo
  {journal} {J. Chem. Phys.}\ }\textbf {\bibinfo {volume} {132}},\ \bibinfo
  {pages} {194102} (\bibinfo {year} {2010})}\BibitemShut {NoStop}%
\bibitem [{\citenamefont {Jabbari-Farouji}\ and\ \citenamefont
  {Trizac}(2012)}]{Jabbari-FaroujiTrizac12}%
  \BibitemOpen
  \bibfield  {author} {\bibinfo {author} {\bibfnamefont {S.}~\bibnamefont
  {Jabbari-Farouji}}\ and\ \bibinfo {author} {\bibfnamefont {E.}~\bibnamefont
  {Trizac}},\ }\bibfield  {title} {\bibinfo {title} {Dynamic monte carlo
  simulations of anisotropic colloids},\ }\href@noop {} {\bibfield  {journal}
  {\bibinfo  {journal} {J. Chem. Phys.}\ }\textbf {\bibinfo {volume} {137}},\
  \bibinfo {pages} {054107} (\bibinfo {year} {2012})}\BibitemShut {NoStop}%
\bibitem [{\citenamefont {Newman}\ and\ \citenamefont
  {Barkema}(1999)}]{MCBookOfNewmanBarkema}%
  \BibitemOpen
  \bibfield  {author} {\bibinfo {author} {\bibfnamefont {M.~E.~J.}\
  \bibnamefont {Newman}}\ and\ \bibinfo {author} {\bibfnamefont {G.~T.}\
  \bibnamefont {Barkema}},\ }\href@noop {} {\emph {\bibinfo {title} {Monte
  Carlo Methods in Statistical Physics}}}\ (\bibinfo  {publisher} {Oxford
  University Press},\ \bibinfo {year} {1999})\BibitemShut {NoStop}%
\bibitem [{\citenamefont {Peruani}\ \emph {et~al.}(2011)\citenamefont
  {Peruani}, \citenamefont {Klauss}, \citenamefont {Deutsch},\ and\
  \citenamefont {Voss-Boehme}}]{peruani2011traffic}%
  \BibitemOpen
  \bibfield  {author} {\bibinfo {author} {\bibfnamefont {F.}~\bibnamefont
  {Peruani}}, \bibinfo {author} {\bibfnamefont {T.}~\bibnamefont {Klauss}},
  \bibinfo {author} {\bibfnamefont {A.}~\bibnamefont {Deutsch}},\ and\ \bibinfo
  {author} {\bibfnamefont {A.}~\bibnamefont {Voss-Boehme}},\ }\bibfield
  {title} {\bibinfo {title} {Traffic jams, gliders, and bands in the quest for
  collective motion of self-propelled particles},\ }\href@noop {} {\bibfield
  {journal} {\bibinfo  {journal} {Phys. Rev. Lett.}\ }\textbf {\bibinfo
  {volume} {106}},\ \bibinfo {pages} {128101} (\bibinfo {year}
  {2011})}\BibitemShut {NoStop}%
\bibitem [{\citenamefont {Thompson}\ \emph {et~al.}(2011)\citenamefont
  {Thompson}, \citenamefont {Tailleur}, \citenamefont {Cates},\ and\
  \citenamefont {Blythe}}]{thompson2011lattice}%
  \BibitemOpen
  \bibfield  {author} {\bibinfo {author} {\bibfnamefont {A.~G.}\ \bibnamefont
  {Thompson}}, \bibinfo {author} {\bibfnamefont {J.}~\bibnamefont {Tailleur}},
  \bibinfo {author} {\bibfnamefont {M.~E.}\ \bibnamefont {Cates}},\ and\
  \bibinfo {author} {\bibfnamefont {R.~A.}\ \bibnamefont {Blythe}},\ }\bibfield
   {title} {\bibinfo {title} {Lattice models of nonequilibrium bacterial
  dynamics},\ }\href@noop {} {\bibfield  {journal} {\bibinfo  {journal} {J.
  Stat. Mech. Theor. Exp.}\ }\textbf {\bibinfo {volume} {2011}},\ \bibinfo
  {pages} {P02029} (\bibinfo {year} {2011})}\BibitemShut {NoStop}%
\bibitem [{\citenamefont {Soto}\ and\ \citenamefont
  {Golestanian}(2014)}]{soto2014self}%
  \BibitemOpen
  \bibfield  {author} {\bibinfo {author} {\bibfnamefont {R.}~\bibnamefont
  {Soto}}\ and\ \bibinfo {author} {\bibfnamefont {R.}~\bibnamefont
  {Golestanian}},\ }\bibfield  {title} {\bibinfo {title} {Self-assembly of
  catalytically active colloidal molecules: Tailoring activity through surface
  chemistry},\ }\href@noop {} {\bibfield  {journal} {\bibinfo  {journal} {Phys.
  Rev. Lett.}\ }\textbf {\bibinfo {volume} {112}},\ \bibinfo {pages} {068301}
  (\bibinfo {year} {2014})}\BibitemShut {NoStop}%
\bibitem [{\citenamefont {Levis}\ and\ \citenamefont
  {Berthier}(2014)}]{LevisBerthierKMC2014}%
  \BibitemOpen
  \bibfield  {author} {\bibinfo {author} {\bibfnamefont {D.}~\bibnamefont
  {Levis}}\ and\ \bibinfo {author} {\bibfnamefont {L.}~\bibnamefont
  {Berthier}},\ }\bibfield  {title} {\bibinfo {title} {Clustering and
  heterogeneous dynamics in a kinetic monte carlo model of self-propelled hard
  disks},\ }\href@noop {} {\bibfield  {journal} {\bibinfo  {journal} {Phys.
  Rev. E}\ }\textbf {\bibinfo {volume} {89}},\ \bibinfo {pages} {062301}
  (\bibinfo {year} {2014})}\BibitemShut {NoStop}%
\bibitem [{\citenamefont {Berthier}(2014)}]{BerthierAKMC2014}%
  \BibitemOpen
  \bibfield  {author} {\bibinfo {author} {\bibfnamefont {L.}~\bibnamefont
  {Berthier}},\ }\bibfield  {title} {\bibinfo {title} {Nonequilibrium glassy
  dynamics of self-propelled hard disks},\ }\href@noop {} {\bibfield  {journal}
  {\bibinfo  {journal} {Phys. Rev. Lett.}\ }\textbf {\bibinfo {volume} {112}},\
  \bibinfo {pages} {220602} (\bibinfo {year} {2014})}\BibitemShut {NoStop}%
\bibitem [{\citenamefont {Levis}\ and\ \citenamefont
  {Berthier}(2015)}]{Levis_2015}%
  \BibitemOpen
  \bibfield  {author} {\bibinfo {author} {\bibfnamefont {D.}~\bibnamefont
  {Levis}}\ and\ \bibinfo {author} {\bibfnamefont {L.}~\bibnamefont
  {Berthier}},\ }\bibfield  {title} {\bibinfo {title} {From single-particle to
  collective effective temperatures in an active fluid of self-propelled
  particles},\ }\href {https://doi.org/10.1209/0295-5075/111/60006} {\bibfield
  {journal} {\bibinfo  {journal} {{EPL}}\ }\textbf {\bibinfo {volume} {111}},\
  \bibinfo {pages} {60006} (\bibinfo {year} {2015})}\BibitemShut {NoStop}%
\bibitem [{\citenamefont {Sep{\'u}lveda}\ and\ \citenamefont
  {Soto}(2016)}]{sepulveda2016coarsening}%
  \BibitemOpen
  \bibfield  {author} {\bibinfo {author} {\bibfnamefont {N.}~\bibnamefont
  {Sep{\'u}lveda}}\ and\ \bibinfo {author} {\bibfnamefont {R.}~\bibnamefont
  {Soto}},\ }\bibfield  {title} {\bibinfo {title} {Coarsening and clustering in
  run-and-tumble dynamics with short-range exclusion},\ }\href@noop {}
  {\bibfield  {journal} {\bibinfo  {journal} {Phys. Rev. E}\ }\textbf {\bibinfo
  {volume} {94}},\ \bibinfo {pages} {022603} (\bibinfo {year}
  {2016})}\BibitemShut {NoStop}%
\bibitem [{\citenamefont {Manacorda}\ and\ \citenamefont
  {Puglisi}(2017)}]{manacorda2017lattice}%
  \BibitemOpen
  \bibfield  {author} {\bibinfo {author} {\bibfnamefont {A.}~\bibnamefont
  {Manacorda}}\ and\ \bibinfo {author} {\bibfnamefont {A.}~\bibnamefont
  {Puglisi}},\ }\bibfield  {title} {\bibinfo {title} {Lattice model to derive
  the fluctuating hydrodynamics of active particles with inertia},\ }\href@noop
  {} {\bibfield  {journal} {\bibinfo  {journal} {Phys. Rev. Lett.}\ }\textbf
  {\bibinfo {volume} {119}},\ \bibinfo {pages} {208003} (\bibinfo {year}
  {2017})}\BibitemShut {NoStop}%
\bibitem [{\citenamefont {Klamser}\ \emph {et~al.}(2018)\citenamefont
  {Klamser}, \citenamefont {Kapfer},\ and\ \citenamefont
  {Krauth}}]{KKK_AKMC2018}%
  \BibitemOpen
  \bibfield  {author} {\bibinfo {author} {\bibfnamefont {J.~U.}\ \bibnamefont
  {Klamser}}, \bibinfo {author} {\bibfnamefont {S.~C.}\ \bibnamefont
  {Kapfer}},\ and\ \bibinfo {author} {\bibfnamefont {W.}~\bibnamefont
  {Krauth}},\ }\bibfield  {title} {\bibinfo {title} {Thermodynamic phases in
  two-dimensional active matter},\ }\href@noop {} {\bibfield  {journal}
  {\bibinfo  {journal} {Nat Commun.}\ }\textbf {\bibinfo {volume} {9}},\
  \bibinfo {pages} {5045} (\bibinfo {year} {2018})}\BibitemShut {NoStop}%
\bibitem [{\citenamefont {Whitelam}\ \emph {et~al.}(2018)\citenamefont
  {Whitelam}, \citenamefont {Klymko},\ and\ \citenamefont
  {Mandal}}]{whitelam2018phase}%
  \BibitemOpen
  \bibfield  {author} {\bibinfo {author} {\bibfnamefont {S.}~\bibnamefont
  {Whitelam}}, \bibinfo {author} {\bibfnamefont {K.}~\bibnamefont {Klymko}},\
  and\ \bibinfo {author} {\bibfnamefont {D.}~\bibnamefont {Mandal}},\
  }\bibfield  {title} {\bibinfo {title} {Phase separation and large deviations
  of lattice active matter},\ }\href@noop {} {\bibfield  {journal} {\bibinfo
  {journal} {J. Chem. Phys.}\ }\textbf {\bibinfo {volume} {148}},\ \bibinfo
  {pages} {154902} (\bibinfo {year} {2018})}\BibitemShut {NoStop}%
\bibitem [{\citenamefont {Kourbane-Houssene}\ \emph {et~al.}(2018)\citenamefont
  {Kourbane-Houssene}, \citenamefont {Erignoux}, \citenamefont {Bodineau},\
  and\ \citenamefont {Tailleur}}]{kourbane2018exact}%
  \BibitemOpen
  \bibfield  {author} {\bibinfo {author} {\bibfnamefont {M.}~\bibnamefont
  {Kourbane-Houssene}}, \bibinfo {author} {\bibfnamefont {C.}~\bibnamefont
  {Erignoux}}, \bibinfo {author} {\bibfnamefont {T.}~\bibnamefont {Bodineau}},\
  and\ \bibinfo {author} {\bibfnamefont {J.}~\bibnamefont {Tailleur}},\
  }\bibfield  {title} {\bibinfo {title} {Exact hydrodynamic description of
  active lattice gases},\ }\href@noop {} {\bibfield  {journal} {\bibinfo
  {journal} {Phys. Rev. Lett.}\ }\textbf {\bibinfo {volume} {120}},\ \bibinfo
  {pages} {268003} (\bibinfo {year} {2018})}\BibitemShut {NoStop}%
\bibitem [{\citenamefont {Klamser}\ \emph {et~al.}(2019)\citenamefont
  {Klamser}, \citenamefont {Kapfer},\ and\ \citenamefont
  {Krauth}}]{KKK_AKMC2019}%
  \BibitemOpen
  \bibfield  {author} {\bibinfo {author} {\bibfnamefont {J.~U.}\ \bibnamefont
  {Klamser}}, \bibinfo {author} {\bibfnamefont {S.~C.}\ \bibnamefont
  {Kapfer}},\ and\ \bibinfo {author} {\bibfnamefont {W.}~\bibnamefont
  {Krauth}},\ }\bibfield  {title} {\bibinfo {title} {A kinetic-monte carlo
  perspective on active matter},\ }\href@noop {} {\bibfield  {journal}
  {\bibinfo  {journal} {J. Chem. Phys.}\ }\textbf {\bibinfo {volume} {150}},\
  \bibinfo {pages} {144113} (\bibinfo {year} {2019})}\BibitemShut {NoStop}%
\bibitem [{\citenamefont {Shi}\ \emph {et~al.}(2020)\citenamefont {Shi},
  \citenamefont {Fausti}, \citenamefont {Chat{\'e}}, \citenamefont {Nardini},\
  and\ \citenamefont {Solon}}]{shi2020self}%
  \BibitemOpen
  \bibfield  {author} {\bibinfo {author} {\bibfnamefont {X.-q.}\ \bibnamefont
  {Shi}}, \bibinfo {author} {\bibfnamefont {G.}~\bibnamefont {Fausti}},
  \bibinfo {author} {\bibfnamefont {H.}~\bibnamefont {Chat{\'e}}}, \bibinfo
  {author} {\bibfnamefont {C.}~\bibnamefont {Nardini}},\ and\ \bibinfo {author}
  {\bibfnamefont {A.}~\bibnamefont {Solon}},\ }\bibfield  {title} {\bibinfo
  {title} {Self-organized critical coexistence phase in repulsive active
  particles},\ }\href@noop {} {\bibfield  {journal} {\bibinfo  {journal} {Phys.
  Rev. Lett.}\ }\textbf {\bibinfo {volume} {125}},\ \bibinfo {pages} {168001}
  (\bibinfo {year} {2020})}\BibitemShut {NoStop}%
\bibitem [{\citenamefont {Ro}\ \emph {et~al.}(2021)\citenamefont {Ro},
  \citenamefont {Kafri}, \citenamefont {Kardar},\ and\ \citenamefont
  {Tailleur}}]{ro2021disorder}%
  \BibitemOpen
  \bibfield  {author} {\bibinfo {author} {\bibfnamefont {S.}~\bibnamefont
  {Ro}}, \bibinfo {author} {\bibfnamefont {Y.}~\bibnamefont {Kafri}}, \bibinfo
  {author} {\bibfnamefont {M.}~\bibnamefont {Kardar}},\ and\ \bibinfo {author}
  {\bibfnamefont {J.}~\bibnamefont {Tailleur}},\ }\bibfield  {title} {\bibinfo
  {title} {Disorder-induced long-ranged correlations in scalar active matter},\
  }\href@noop {} {\bibfield  {journal} {\bibinfo  {journal} {Phys. Rev. Lett.}\
  }\textbf {\bibinfo {volume} {126}},\ \bibinfo {pages} {048003} (\bibinfo
  {year} {2021})}\BibitemShut {NoStop}%
\bibitem [{\citenamefont {Marchetti}\ \emph {et~al.}(2013)\citenamefont
  {Marchetti}, \citenamefont {Joanny}, \citenamefont {Ramaswamy}, \citenamefont
  {Liverpool}, \citenamefont {Prost}, \citenamefont {Rao},\ and\ \citenamefont
  {Simha}}]{marchetti2013hydrodynamics}%
  \BibitemOpen
  \bibfield  {author} {\bibinfo {author} {\bibfnamefont {M.~C.}\ \bibnamefont
  {Marchetti}}, \bibinfo {author} {\bibfnamefont {J.-F.}\ \bibnamefont
  {Joanny}}, \bibinfo {author} {\bibfnamefont {S.}~\bibnamefont {Ramaswamy}},
  \bibinfo {author} {\bibfnamefont {T.~B.}\ \bibnamefont {Liverpool}}, \bibinfo
  {author} {\bibfnamefont {J.}~\bibnamefont {Prost}}, \bibinfo {author}
  {\bibfnamefont {M.}~\bibnamefont {Rao}},\ and\ \bibinfo {author}
  {\bibfnamefont {R.~A.}\ \bibnamefont {Simha}},\ }\bibfield  {title} {\bibinfo
  {title} {Hydrodynamics of soft active matter},\ }\href@noop {} {\bibfield
  {journal} {\bibinfo  {journal} {Rev. Mod. Phys.}\ }\textbf {\bibinfo {volume}
  {85}},\ \bibinfo {pages} {1143} (\bibinfo {year} {2013})}\BibitemShut
  {NoStop}%
\bibitem [{\citenamefont {Bechinger}\ \emph {et~al.}(2016)\citenamefont
  {Bechinger}, \citenamefont {Di~Leonardo}, \citenamefont {L{\"o}wen},
  \citenamefont {Reichhardt}, \citenamefont {Volpe},\ and\ \citenamefont
  {Volpe}}]{bechinger2016active}%
  \BibitemOpen
  \bibfield  {author} {\bibinfo {author} {\bibfnamefont {C.}~\bibnamefont
  {Bechinger}}, \bibinfo {author} {\bibfnamefont {R.}~\bibnamefont
  {Di~Leonardo}}, \bibinfo {author} {\bibfnamefont {H.}~\bibnamefont
  {L{\"o}wen}}, \bibinfo {author} {\bibfnamefont {C.}~\bibnamefont
  {Reichhardt}}, \bibinfo {author} {\bibfnamefont {G.}~\bibnamefont {Volpe}},\
  and\ \bibinfo {author} {\bibfnamefont {G.}~\bibnamefont {Volpe}},\ }\bibfield
   {title} {\bibinfo {title} {Active particles in complex and crowded
  environments},\ }\href@noop {} {\bibfield  {journal} {\bibinfo  {journal}
  {Rev. Mod. Phys.}\ }\textbf {\bibinfo {volume} {88}},\ \bibinfo {pages}
  {045006} (\bibinfo {year} {2016})}\BibitemShut {NoStop}%
\bibitem [{\citenamefont {Romanczuk}\ \emph {et~al.}(2012)\citenamefont
  {Romanczuk}, \citenamefont {B{\"a}r}, \citenamefont {Ebeling}, \citenamefont
  {Lindner},\ and\ \citenamefont {Schimansky-Geier}}]{romanczuk2012active}%
  \BibitemOpen
  \bibfield  {author} {\bibinfo {author} {\bibfnamefont {P.}~\bibnamefont
  {Romanczuk}}, \bibinfo {author} {\bibfnamefont {M.}~\bibnamefont {B{\"a}r}},
  \bibinfo {author} {\bibfnamefont {W.}~\bibnamefont {Ebeling}}, \bibinfo
  {author} {\bibfnamefont {B.}~\bibnamefont {Lindner}},\ and\ \bibinfo {author}
  {\bibfnamefont {L.}~\bibnamefont {Schimansky-Geier}},\ }\bibfield  {title}
  {\bibinfo {title} {Active brownian particles},\ }\href@noop {} {\bibfield
  {journal} {\bibinfo  {journal} {The European Physical Journal Special
  Topics}\ }\textbf {\bibinfo {volume} {202}},\ \bibinfo {pages} {1} (\bibinfo
  {year} {2012})}\BibitemShut {NoStop}%
\bibitem [{\citenamefont {Vicsek}\ and\ \citenamefont
  {Zafeiris}(2012)}]{vicsek2012collective}%
  \BibitemOpen
  \bibfield  {author} {\bibinfo {author} {\bibfnamefont {T.}~\bibnamefont
  {Vicsek}}\ and\ \bibinfo {author} {\bibfnamefont {A.}~\bibnamefont
  {Zafeiris}},\ }\bibfield  {title} {\bibinfo {title} {Collective motion},\
  }\href@noop {} {\bibfield  {journal} {\bibinfo  {journal} {Phys. Rep.}\
  }\textbf {\bibinfo {volume} {517}},\ \bibinfo {pages} {71} (\bibinfo {year}
  {2012})}\BibitemShut {NoStop}%
\bibitem [{\citenamefont {Chat{\'e}}(2020)}]{chate2020dry}%
  \BibitemOpen
  \bibfield  {author} {\bibinfo {author} {\bibfnamefont {H.}~\bibnamefont
  {Chat{\'e}}},\ }\bibfield  {title} {\bibinfo {title} {Dry aligning dilute
  active matter},\ }\href@noop {} {\bibfield  {journal} {\bibinfo  {journal}
  {Annual Review of Condensed Matter Physics}\ }\textbf {\bibinfo {volume}
  {11}},\ \bibinfo {pages} {189} (\bibinfo {year} {2020})}\BibitemShut
  {NoStop}%
\bibitem [{\citenamefont {Cates}\ and\ \citenamefont
  {Tailleur}(2015)}]{cates2015motility}%
  \BibitemOpen
  \bibfield  {author} {\bibinfo {author} {\bibfnamefont {M.~E.}\ \bibnamefont
  {Cates}}\ and\ \bibinfo {author} {\bibfnamefont {J.}~\bibnamefont
  {Tailleur}},\ }\bibfield  {title} {\bibinfo {title} {Motility-induced phase
  separation},\ }\href@noop {} {\bibfield  {journal} {\bibinfo  {journal}
  {Annu. Rev. Condens. Matter Phys.}\ }\textbf {\bibinfo {volume} {6}},\
  \bibinfo {pages} {219} (\bibinfo {year} {2015})}\BibitemShut {NoStop}%
\bibitem [{\citenamefont {Wensink}\ \emph {et~al.}(2012)\citenamefont
  {Wensink}, \citenamefont {Dunkel}, \citenamefont {Heidenreich}, \citenamefont
  {Drescher}, \citenamefont {Goldstein}, \citenamefont {L{\"o}wen},\ and\
  \citenamefont {Yeomans}}]{wensink2012meso}%
  \BibitemOpen
  \bibfield  {author} {\bibinfo {author} {\bibfnamefont {H.~H.}\ \bibnamefont
  {Wensink}}, \bibinfo {author} {\bibfnamefont {J.}~\bibnamefont {Dunkel}},
  \bibinfo {author} {\bibfnamefont {S.}~\bibnamefont {Heidenreich}}, \bibinfo
  {author} {\bibfnamefont {K.}~\bibnamefont {Drescher}}, \bibinfo {author}
  {\bibfnamefont {R.~E.}\ \bibnamefont {Goldstein}}, \bibinfo {author}
  {\bibfnamefont {H.}~\bibnamefont {L{\"o}wen}},\ and\ \bibinfo {author}
  {\bibfnamefont {J.~M.}\ \bibnamefont {Yeomans}},\ }\bibfield  {title}
  {\bibinfo {title} {Meso-scale turbulence in living fluids},\ }\href@noop {}
  {\bibfield  {journal} {\bibinfo  {journal} {Proceedings of the National
  Academy of Sciences}\ }\textbf {\bibinfo {volume} {109}},\ \bibinfo {pages}
  {14308} (\bibinfo {year} {2012})}\BibitemShut {NoStop}%
\bibitem [{\citenamefont {Stenhammar}\ \emph {et~al.}(2017)\citenamefont
  {Stenhammar}, \citenamefont {Nardini}, \citenamefont {Nash}, \citenamefont
  {Marenduzzo},\ and\ \citenamefont {Morozov}}]{stenhammar2017role}%
  \BibitemOpen
  \bibfield  {author} {\bibinfo {author} {\bibfnamefont {J.}~\bibnamefont
  {Stenhammar}}, \bibinfo {author} {\bibfnamefont {C.}~\bibnamefont {Nardini}},
  \bibinfo {author} {\bibfnamefont {R.~W.}\ \bibnamefont {Nash}}, \bibinfo
  {author} {\bibfnamefont {D.}~\bibnamefont {Marenduzzo}},\ and\ \bibinfo
  {author} {\bibfnamefont {A.}~\bibnamefont {Morozov}},\ }\bibfield  {title}
  {\bibinfo {title} {Role of correlations in the collective behavior of
  microswimmer suspensions},\ }\href@noop {} {\bibfield  {journal} {\bibinfo
  {journal} {Phys. Rev. Lett.}\ }\textbf {\bibinfo {volume} {119}},\ \bibinfo
  {pages} {028005} (\bibinfo {year} {2017})}\BibitemShut {NoStop}%
\bibitem [{\citenamefont {Wu}\ \emph {et~al.}(2017)\citenamefont {Wu},
  \citenamefont {Hishamunda}, \citenamefont {Chen}, \citenamefont {DeCamp},
  \citenamefont {Chang}, \citenamefont {Fern{\'a}ndez-Nieves}, \citenamefont
  {Fraden},\ and\ \citenamefont {Dogic}}]{wu2017transition}%
  \BibitemOpen
  \bibfield  {author} {\bibinfo {author} {\bibfnamefont {K.-T.}\ \bibnamefont
  {Wu}}, \bibinfo {author} {\bibfnamefont {J.~B.}\ \bibnamefont {Hishamunda}},
  \bibinfo {author} {\bibfnamefont {D.~T.}\ \bibnamefont {Chen}}, \bibinfo
  {author} {\bibfnamefont {S.~J.}\ \bibnamefont {DeCamp}}, \bibinfo {author}
  {\bibfnamefont {Y.-W.}\ \bibnamefont {Chang}}, \bibinfo {author}
  {\bibfnamefont {A.}~\bibnamefont {Fern{\'a}ndez-Nieves}}, \bibinfo {author}
  {\bibfnamefont {S.}~\bibnamefont {Fraden}},\ and\ \bibinfo {author}
  {\bibfnamefont {Z.}~\bibnamefont {Dogic}},\ }\bibfield  {title} {\bibinfo
  {title} {Transition from turbulent to coherent flows in confined
  three-dimensional active fluids},\ }\href@noop {} {\ \textbf {\bibinfo
  {volume} {355}},\ \bibinfo {pages} {eaal1979} (\bibinfo {year}
  {2017})}\BibitemShut {NoStop}%
\bibitem [{\citenamefont {Gr{\'e}goire}\ and\ \citenamefont
  {Chat{\'e}}(2004)}]{gregoire2004onset}%
  \BibitemOpen
  \bibfield  {author} {\bibinfo {author} {\bibfnamefont {G.}~\bibnamefont
  {Gr{\'e}goire}}\ and\ \bibinfo {author} {\bibfnamefont {H.}~\bibnamefont
  {Chat{\'e}}},\ }\bibfield  {title} {\bibinfo {title} {Onset of collective and
  cohesive motion},\ }\href@noop {} {\bibfield  {journal} {\bibinfo  {journal}
  {Phys. Rev. Lett.}\ }\textbf {\bibinfo {volume} {92}},\ \bibinfo {pages}
  {025702} (\bibinfo {year} {2004})}\BibitemShut {NoStop}%
\bibitem [{\citenamefont {Weber}\ \emph {et~al.}(2013)\citenamefont {Weber},
  \citenamefont {Hanke}, \citenamefont {Deseigne}, \citenamefont {L{\'e}onard},
  \citenamefont {Dauchot}, \citenamefont {Frey},\ and\ \citenamefont
  {Chat{\'e}}}]{weber2013long}%
  \BibitemOpen
  \bibfield  {author} {\bibinfo {author} {\bibfnamefont {C.~A.}\ \bibnamefont
  {Weber}}, \bibinfo {author} {\bibfnamefont {T.}~\bibnamefont {Hanke}},
  \bibinfo {author} {\bibfnamefont {J.}~\bibnamefont {Deseigne}}, \bibinfo
  {author} {\bibfnamefont {S.}~\bibnamefont {L{\'e}onard}}, \bibinfo {author}
  {\bibfnamefont {O.}~\bibnamefont {Dauchot}}, \bibinfo {author} {\bibfnamefont
  {E.}~\bibnamefont {Frey}},\ and\ \bibinfo {author} {\bibfnamefont
  {H.}~\bibnamefont {Chat{\'e}}},\ }\bibfield  {title} {\bibinfo {title}
  {Long-range ordering of vibrated polar disks},\ }\href@noop {} {\bibfield
  {journal} {\bibinfo  {journal} {Phys. Rev. Lett.}\ }\textbf {\bibinfo
  {volume} {110}},\ \bibinfo {pages} {208001} (\bibinfo {year}
  {2013})}\BibitemShut {NoStop}%
\bibitem [{\citenamefont {Solon}\ and\ \citenamefont
  {Tailleur}(2013)}]{solon2013revisiting}%
  \BibitemOpen
  \bibfield  {author} {\bibinfo {author} {\bibfnamefont {A.}~\bibnamefont
  {Solon}}\ and\ \bibinfo {author} {\bibfnamefont {J.}~\bibnamefont
  {Tailleur}},\ }\bibfield  {title} {\bibinfo {title} {Revisiting the flocking
  transition using active spins},\ }\href@noop {} {\bibfield  {journal}
  {\bibinfo  {journal} {Phys. Rev. Lett.}\ }\textbf {\bibinfo {volume} {111}},\
  \bibinfo {pages} {078101} (\bibinfo {year} {2013})}\BibitemShut {NoStop}%
\bibitem [{\citenamefont {Solon}\ \emph
  {et~al.}(2015{\natexlab{a}})\citenamefont {Solon}, \citenamefont
  {Chat{\'e}},\ and\ \citenamefont {Tailleur}}]{solon2015phase}%
  \BibitemOpen
  \bibfield  {author} {\bibinfo {author} {\bibfnamefont {A.~P.}\ \bibnamefont
  {Solon}}, \bibinfo {author} {\bibfnamefont {H.}~\bibnamefont {Chat{\'e}}},\
  and\ \bibinfo {author} {\bibfnamefont {J.}~\bibnamefont {Tailleur}},\
  }\bibfield  {title} {\bibinfo {title} {From phase to microphase separation in
  flocking models: The essential role of nonequilibrium fluctuations},\
  }\href@noop {} {\bibfield  {journal} {\bibinfo  {journal} {Phys. Rev. Lett.}\
  }\textbf {\bibinfo {volume} {114}},\ \bibinfo {pages} {068101} (\bibinfo
  {year} {2015}{\natexlab{a}})}\BibitemShut {NoStop}%
\bibitem [{\citenamefont {Digregorio}\ \emph {et~al.}(2018)\citenamefont
  {Digregorio}, \citenamefont {Levis}, \citenamefont {Suma}, \citenamefont
  {Cugliandolo}, \citenamefont {Gonnella},\ and\ \citenamefont
  {Pagonabarraga}}]{MeltingABP2018}%
  \BibitemOpen
  \bibfield  {author} {\bibinfo {author} {\bibfnamefont {P.}~\bibnamefont
  {Digregorio}}, \bibinfo {author} {\bibfnamefont {D.}~\bibnamefont {Levis}},
  \bibinfo {author} {\bibfnamefont {A.}~\bibnamefont {Suma}}, \bibinfo {author}
  {\bibfnamefont {L.~F.}\ \bibnamefont {Cugliandolo}}, \bibinfo {author}
  {\bibfnamefont {G.}~\bibnamefont {Gonnella}},\ and\ \bibinfo {author}
  {\bibfnamefont {I.}~\bibnamefont {Pagonabarraga}},\ }\bibfield  {title}
  {\bibinfo {title} {Full phase diagram of active brownian disks: From melting
  to motility-induced phase separation},\ }\href@noop {} {\bibfield  {journal}
  {\bibinfo  {journal} {Phys. Rev. Lett.}\ }\textbf {\bibinfo {volume} {121}},\
  \bibinfo {pages} {098003} (\bibinfo {year} {2018})}\BibitemShut {NoStop}%
\bibitem [{\citenamefont {Schnitzer}(1993)}]{schnitzer1993theory}%
  \BibitemOpen
  \bibfield  {author} {\bibinfo {author} {\bibfnamefont {M.~J.}\ \bibnamefont
  {Schnitzer}},\ }\bibfield  {title} {\bibinfo {title} {Theory of continuum
  random walks and application to chemotaxis},\ }\href@noop {} {\bibfield
  {journal} {\bibinfo  {journal} {Physical Review E}\ }\textbf {\bibinfo
  {volume} {48}},\ \bibinfo {pages} {2553} (\bibinfo {year}
  {1993})}\BibitemShut {NoStop}%
\bibitem [{\citenamefont {Berg}(2008)}]{berg2008coli}%
  \BibitemOpen
  \bibfield  {author} {\bibinfo {author} {\bibfnamefont {H.~C.}\ \bibnamefont
  {Berg}},\ }\href@noop {} {\emph {\bibinfo {title} {E. coli in Motion}}}\
  (\bibinfo  {publisher} {Springer Science \& Business Media},\ \bibinfo {year}
  {2008})\BibitemShut {NoStop}%
\bibitem [{\citenamefont {Fily}\ and\ \citenamefont
  {Marchetti}(2012)}]{fily2012athermal}%
  \BibitemOpen
  \bibfield  {author} {\bibinfo {author} {\bibfnamefont {Y.}~\bibnamefont
  {Fily}}\ and\ \bibinfo {author} {\bibfnamefont {M.~C.}\ \bibnamefont
  {Marchetti}},\ }\bibfield  {title} {\bibinfo {title} {Athermal phase
  separation of self-propelled particles with no alignment},\ }\href@noop {}
  {\bibfield  {journal} {\bibinfo  {journal} {Phys. Rev. Lett.}\ }\textbf
  {\bibinfo {volume} {108}},\ \bibinfo {pages} {235702} (\bibinfo {year}
  {2012})}\BibitemShut {NoStop}%
\bibitem [{\citenamefont {Farage}\ \emph {et~al.}(2015)\citenamefont {Farage},
  \citenamefont {Krinninger},\ and\ \citenamefont
  {Brader}}]{farage2015effective}%
  \BibitemOpen
  \bibfield  {author} {\bibinfo {author} {\bibfnamefont {T.~F.}\ \bibnamefont
  {Farage}}, \bibinfo {author} {\bibfnamefont {P.}~\bibnamefont {Krinninger}},\
  and\ \bibinfo {author} {\bibfnamefont {J.~M.}\ \bibnamefont {Brader}},\
  }\bibfield  {title} {\bibinfo {title} {Effective interactions in active
  brownian suspensions},\ }\href@noop {} {\bibfield  {journal} {\bibinfo
  {journal} {Physical Review E}\ }\textbf {\bibinfo {volume} {91}},\ \bibinfo
  {pages} {042310} (\bibinfo {year} {2015})}\BibitemShut {NoStop}%
\bibitem [{\citenamefont {Szamel}(2014)}]{szamel2014self}%
  \BibitemOpen
  \bibfield  {author} {\bibinfo {author} {\bibfnamefont {G.}~\bibnamefont
  {Szamel}},\ }\bibfield  {title} {\bibinfo {title} {Self-propelled particle in
  an external potential: Existence of an effective temperature},\ }\href@noop
  {} {\bibfield  {journal} {\bibinfo  {journal} {Physical Review E}\ }\textbf
  {\bibinfo {volume} {90}},\ \bibinfo {pages} {012111} (\bibinfo {year}
  {2014})}\BibitemShut {NoStop}%
\bibitem [{\citenamefont {Martin}\ \emph {et~al.}(2021)\citenamefont {Martin},
  \citenamefont {O'Byrne}, \citenamefont {Cates}, \citenamefont {Fodor},
  \citenamefont {Nardini}, \citenamefont {Tailleur},\ and\ \citenamefont {van
  Wijland}}]{martin2021statistical}%
  \BibitemOpen
  \bibfield  {author} {\bibinfo {author} {\bibfnamefont {D.}~\bibnamefont
  {Martin}}, \bibinfo {author} {\bibfnamefont {J.}~\bibnamefont {O'Byrne}},
  \bibinfo {author} {\bibfnamefont {M.~E.}\ \bibnamefont {Cates}}, \bibinfo
  {author} {\bibfnamefont {{\'E}.}~\bibnamefont {Fodor}}, \bibinfo {author}
  {\bibfnamefont {C.}~\bibnamefont {Nardini}}, \bibinfo {author} {\bibfnamefont
  {J.}~\bibnamefont {Tailleur}},\ and\ \bibinfo {author} {\bibfnamefont
  {F.}~\bibnamefont {van Wijland}},\ }\bibfield  {title} {\bibinfo {title}
  {Statistical mechanics of active ornstein-uhlenbeck particles},\ }\href@noop
  {} {\bibfield  {journal} {\bibinfo  {journal} {Physical Review E}\ }\textbf
  {\bibinfo {volume} {103}},\ \bibinfo {pages} {032607} (\bibinfo {year}
  {2021})}\BibitemShut {NoStop}%
\bibitem [{sup()}]{supp}%
  \BibitemOpen
  \href@noop {} {}\bibinfo {note} {See Supplemental Material [url] which
  includes numerical details.}\BibitemShut {Stop}%
\bibitem [{\citenamefont {Galajda}\ \emph {et~al.}(2007)\citenamefont
  {Galajda}, \citenamefont {Keymer}, \citenamefont {Chaikin},\ and\
  \citenamefont {Austin}}]{galajda2007wall}%
  \BibitemOpen
  \bibfield  {author} {\bibinfo {author} {\bibfnamefont {P.}~\bibnamefont
  {Galajda}}, \bibinfo {author} {\bibfnamefont {J.}~\bibnamefont {Keymer}},
  \bibinfo {author} {\bibfnamefont {P.}~\bibnamefont {Chaikin}},\ and\ \bibinfo
  {author} {\bibfnamefont {R.}~\bibnamefont {Austin}},\ }\bibfield  {title}
  {\bibinfo {title} {A wall of funnels concentrates swimming bacteria},\
  }\href@noop {} {\bibfield  {journal} {\bibinfo  {journal} {Journal of
  bacteriology}\ }\textbf {\bibinfo {volume} {189}},\ \bibinfo {pages} {8704}
  (\bibinfo {year} {2007})}\BibitemShut {NoStop}%
\bibitem [{\citenamefont {Wan}\ \emph {et~al.}(2008)\citenamefont {Wan},
  \citenamefont {Reichhardt}, \citenamefont {Nussinov},\ and\ \citenamefont
  {Reichhardt}}]{wan2008rectification}%
  \BibitemOpen
  \bibfield  {author} {\bibinfo {author} {\bibfnamefont {M.}~\bibnamefont
  {Wan}}, \bibinfo {author} {\bibfnamefont {C.~O.}\ \bibnamefont {Reichhardt}},
  \bibinfo {author} {\bibfnamefont {Z.}~\bibnamefont {Nussinov}},\ and\
  \bibinfo {author} {\bibfnamefont {C.}~\bibnamefont {Reichhardt}},\ }\bibfield
   {title} {\bibinfo {title} {Rectification of swimming bacteria and
  self-driven particle systems by arrays of asymmetric barriers},\ }\href@noop
  {} {\bibfield  {journal} {\bibinfo  {journal} {Phys. Rev. Lett.}\ }\textbf
  {\bibinfo {volume} {101}},\ \bibinfo {pages} {018102} (\bibinfo {year}
  {2008})}\BibitemShut {NoStop}%
\bibitem [{\citenamefont {Tailleur}\ and\ \citenamefont
  {Cates}(2009)}]{tailleur2009sedimentation}%
  \BibitemOpen
  \bibfield  {author} {\bibinfo {author} {\bibfnamefont {J.}~\bibnamefont
  {Tailleur}}\ and\ \bibinfo {author} {\bibfnamefont {M.}~\bibnamefont
  {Cates}},\ }\bibfield  {title} {\bibinfo {title} {Sedimentation, trapping,
  and rectification of dilute bacteria},\ }\href@noop {} {\bibfield  {journal}
  {\bibinfo  {journal} {EPL (Europhysics Letters)}\ }\textbf {\bibinfo {volume}
  {86}},\ \bibinfo {pages} {60002} (\bibinfo {year} {2009})}\BibitemShut
  {NoStop}%
\bibitem [{\citenamefont {Solon}\ \emph
  {et~al.}(2015{\natexlab{b}})\citenamefont {Solon}, \citenamefont {Fily},
  \citenamefont {Baskaran}, \citenamefont {Cates}, \citenamefont {Kafri},
  \citenamefont {Kardar},\ and\ \citenamefont
  {Tailleur}}]{GenActPresSolon2015}%
  \BibitemOpen
  \bibfield  {author} {\bibinfo {author} {\bibfnamefont {A.~P.}\ \bibnamefont
  {Solon}}, \bibinfo {author} {\bibfnamefont {Y.}~\bibnamefont {Fily}},
  \bibinfo {author} {\bibfnamefont {A.}~\bibnamefont {Baskaran}}, \bibinfo
  {author} {\bibfnamefont {M.~E.}\ \bibnamefont {Cates}}, \bibinfo {author}
  {\bibfnamefont {Y.}~\bibnamefont {Kafri}}, \bibinfo {author} {\bibfnamefont
  {M.}~\bibnamefont {Kardar}},\ and\ \bibinfo {author} {\bibfnamefont
  {J.}~\bibnamefont {Tailleur}},\ }\bibfield  {title} {\bibinfo {title}
  {Pressure is not a state function for generic active fluids},\ }\href@noop {}
  {\bibfield  {journal} {\bibinfo  {journal} {Nat. Phys.}\ }\textbf {\bibinfo
  {volume} {11}},\ \bibinfo {pages} {673} (\bibinfo {year}
  {2015}{\natexlab{b}})}\BibitemShut {NoStop}%
\bibitem [{\citenamefont {Gardiner}\ \emph {et~al.}(1985)\citenamefont
  {Gardiner} \emph {et~al.}}]{gardiner1985handbook}%
  \BibitemOpen
  \bibfield  {author} {\bibinfo {author} {\bibfnamefont {C.~W.}\ \bibnamefont
  {Gardiner}} \emph {et~al.},\ }\href@noop {} {\emph {\bibinfo {title}
  {Handbook of stochastic methods}}},\ Vol.~\bibinfo {volume} {3}\ (\bibinfo
  {publisher} {springer Berlin},\ \bibinfo {year} {1985})\BibitemShut {NoStop}%
\bibitem [{\citenamefont {Van~Kampen}(1992)}]{van1992stochastic}%
  \BibitemOpen
  \bibfield  {author} {\bibinfo {author} {\bibfnamefont {N.~G.}\ \bibnamefont
  {Van~Kampen}},\ }\href@noop {} {\emph {\bibinfo {title} {Stochastic processes
  in physics and chemistry}}},\ Vol.~\bibinfo {volume} {1}\ (\bibinfo
  {publisher} {Elsevier},\ \bibinfo {year} {1992})\BibitemShut {NoStop}%
\bibitem [{\citenamefont {Kikuchi}\ \emph {et~al.}(1991)\citenamefont
  {Kikuchi}, \citenamefont {Yoshida}, \citenamefont {Maekawa},\ and\
  \citenamefont {Watanabe}}]{KIKUCHI1991335}%
  \BibitemOpen
  \bibfield  {author} {\bibinfo {author} {\bibfnamefont {K.}~\bibnamefont
  {Kikuchi}}, \bibinfo {author} {\bibfnamefont {M.}~\bibnamefont {Yoshida}},
  \bibinfo {author} {\bibfnamefont {T.}~\bibnamefont {Maekawa}},\ and\ \bibinfo
  {author} {\bibfnamefont {H.}~\bibnamefont {Watanabe}},\ }\bibfield  {title}
  {\bibinfo {title} {Metropolis monte carlo method as a numerical technique to
  solve the fokker—planck equation},\ }\href
  {https://doi.org/https://doi.org/10.1016/S0009-2614(91)85070-D} {\bibfield
  {journal} {\bibinfo  {journal} {Chem. Phys. Lett.}\ }\textbf {\bibinfo
  {volume} {185}},\ \bibinfo {pages} {335} (\bibinfo {year}
  {1991})}\BibitemShut {NoStop}%
\bibitem [{\citenamefont {Whitelam}\ \emph {et~al.}(2021)\citenamefont
  {Whitelam}, \citenamefont {Selin}, \citenamefont {Park},\ and\ \citenamefont
  {Tamblyn}}]{neuroevolution2020}%
  \BibitemOpen
  \bibfield  {author} {\bibinfo {author} {\bibfnamefont {S.}~\bibnamefont
  {Whitelam}}, \bibinfo {author} {\bibfnamefont {V.}~\bibnamefont {Selin}},
  \bibinfo {author} {\bibfnamefont {S.-W.}\ \bibnamefont {Park}},\ and\
  \bibinfo {author} {\bibfnamefont {I.}~\bibnamefont {Tamblyn}},\ }\href@noop
  {} {\bibinfo {title} {Correspondence between neuroevolution and gradient
  descent}} (\bibinfo {year} {2021}),\ \Eprint
  {https://arxiv.org/abs/2008.06643} {arXiv:2008.06643 [cs.NE]} \BibitemShut
  {NoStop}%
\bibitem [{Kla()}]{Klamserfuture}%
  \BibitemOpen
  \href@noop {} {}\bibinfo {note} {J. Klamser, O. Dauchot, J. Tailleur, in
  preparation}\BibitemShut {NoStop}%
\bibitem [{\citenamefont {Henkes}\ \emph {et~al.}(2011)\citenamefont {Henkes},
  \citenamefont {Fily},\ and\ \citenamefont {Marchetti}}]{henkes2011active}%
  \BibitemOpen
  \bibfield  {author} {\bibinfo {author} {\bibfnamefont {S.}~\bibnamefont
  {Henkes}}, \bibinfo {author} {\bibfnamefont {Y.}~\bibnamefont {Fily}},\ and\
  \bibinfo {author} {\bibfnamefont {M.~C.}\ \bibnamefont {Marchetti}},\
  }\bibfield  {title} {\bibinfo {title} {Active jamming: Self-propelled soft
  particles at high density},\ }\href@noop {} {\bibfield  {journal} {\bibinfo
  {journal} {Physical Review E}\ }\textbf {\bibinfo {volume} {84}},\ \bibinfo
  {pages} {040301(R)} (\bibinfo {year} {2011})}\BibitemShut {NoStop}%
\bibitem [{\citenamefont {Flenner}\ \emph {et~al.}(2016)\citenamefont
  {Flenner}, \citenamefont {Szamel},\ and\ \citenamefont
  {Berthier}}]{flenner2016nonequilibrium}%
  \BibitemOpen
  \bibfield  {author} {\bibinfo {author} {\bibfnamefont {E.}~\bibnamefont
  {Flenner}}, \bibinfo {author} {\bibfnamefont {G.}~\bibnamefont {Szamel}},\
  and\ \bibinfo {author} {\bibfnamefont {L.}~\bibnamefont {Berthier}},\
  }\bibfield  {title} {\bibinfo {title} {The nonequilibrium glassy dynamics of
  self-propelled particles},\ }\href@noop {} {\bibfield  {journal} {\bibinfo
  {journal} {Soft matter}\ }\textbf {\bibinfo {volume} {12}},\ \bibinfo {pages}
  {7136} (\bibinfo {year} {2016})}\BibitemShut {NoStop}%
\bibitem [{\citenamefont {Berthier}\ \emph
  {et~al.}(2019{\natexlab{b}})\citenamefont {Berthier}, \citenamefont
  {Flenner},\ and\ \citenamefont {Szamel}}]{berthier2019glassy}%
  \BibitemOpen
  \bibfield  {author} {\bibinfo {author} {\bibfnamefont {L.}~\bibnamefont
  {Berthier}}, \bibinfo {author} {\bibfnamefont {E.}~\bibnamefont {Flenner}},\
  and\ \bibinfo {author} {\bibfnamefont {G.}~\bibnamefont {Szamel}},\
  }\bibfield  {title} {\bibinfo {title} {Glassy dynamics in dense systems of
  active particles},\ }\href@noop {} {\bibfield  {journal} {\bibinfo  {journal}
  {The Journal of chemical physics}\ }\textbf {\bibinfo {volume} {150}},\
  \bibinfo {pages} {200901} (\bibinfo {year} {2019}{\natexlab{b}})}\BibitemShut
  {NoStop}%
\bibitem [{\citenamefont {Mandal}\ \emph {et~al.}(2020)\citenamefont {Mandal},
  \citenamefont {Bhuyan}, \citenamefont {Chaudhuri}, \citenamefont {Dasgupta},\
  and\ \citenamefont {Rao}}]{mandal2020extreme}%
  \BibitemOpen
  \bibfield  {author} {\bibinfo {author} {\bibfnamefont {R.}~\bibnamefont
  {Mandal}}, \bibinfo {author} {\bibfnamefont {P.~J.}\ \bibnamefont {Bhuyan}},
  \bibinfo {author} {\bibfnamefont {P.}~\bibnamefont {Chaudhuri}}, \bibinfo
  {author} {\bibfnamefont {C.}~\bibnamefont {Dasgupta}},\ and\ \bibinfo
  {author} {\bibfnamefont {M.}~\bibnamefont {Rao}},\ }\bibfield  {title}
  {\bibinfo {title} {Extreme active matter at high densities},\ }\href@noop {}
  {\bibfield  {journal} {\bibinfo  {journal} {Nat. Commun.}\ }\textbf {\bibinfo
  {volume} {11}},\ \bibinfo {pages} {2581} (\bibinfo {year}
  {2020})}\BibitemShut {NoStop}%
\bibitem [{\citenamefont {Matoz-Fernandez}\ \emph {et~al.}(2017)\citenamefont
  {Matoz-Fernandez}, \citenamefont {Martens}, \citenamefont {Sknepnek},
  \citenamefont {Barrat},\ and\ \citenamefont {Henkes}}]{matoz2017cell}%
  \BibitemOpen
  \bibfield  {author} {\bibinfo {author} {\bibfnamefont {D.}~\bibnamefont
  {Matoz-Fernandez}}, \bibinfo {author} {\bibfnamefont {K.}~\bibnamefont
  {Martens}}, \bibinfo {author} {\bibfnamefont {R.}~\bibnamefont {Sknepnek}},
  \bibinfo {author} {\bibfnamefont {J.}~\bibnamefont {Barrat}},\ and\ \bibinfo
  {author} {\bibfnamefont {S.}~\bibnamefont {Henkes}},\ }\bibfield  {title}
  {\bibinfo {title} {Cell division and death inhibit glassy behaviour of
  confluent tissues},\ }\href@noop {} {\bibfield  {journal} {\bibinfo
  {journal} {Soft matter}\ }\textbf {\bibinfo {volume} {13}},\ \bibinfo {pages}
  {3205} (\bibinfo {year} {2017})}\BibitemShut {NoStop}%
\bibitem [{\citenamefont {Loewe}\ \emph {et~al.}(2020)\citenamefont {Loewe},
  \citenamefont {Chiang}, \citenamefont {Marenduzzo},\ and\ \citenamefont
  {Marchetti}}]{loewe2020solid}%
  \BibitemOpen
  \bibfield  {author} {\bibinfo {author} {\bibfnamefont {B.}~\bibnamefont
  {Loewe}}, \bibinfo {author} {\bibfnamefont {M.}~\bibnamefont {Chiang}},
  \bibinfo {author} {\bibfnamefont {D.}~\bibnamefont {Marenduzzo}},\ and\
  \bibinfo {author} {\bibfnamefont {M.~C.}\ \bibnamefont {Marchetti}},\
  }\bibfield  {title} {\bibinfo {title} {Solid-liquid transition of deformable
  and overlapping active particles},\ }\href@noop {} {\bibfield  {journal}
  {\bibinfo  {journal} {Phys. Rev. Lett.}\ }\textbf {\bibinfo {volume} {125}},\
  \bibinfo {pages} {038003} (\bibinfo {year} {2020})}\BibitemShut {NoStop}%
\bibitem [{\citenamefont {Henkes}\ \emph {et~al.}(2020)\citenamefont {Henkes},
  \citenamefont {Kostanjevec}, \citenamefont {Collinson}, \citenamefont
  {Sknepnek},\ and\ \citenamefont {Bertin}}]{henkes2020dense}%
  \BibitemOpen
  \bibfield  {author} {\bibinfo {author} {\bibfnamefont {S.}~\bibnamefont
  {Henkes}}, \bibinfo {author} {\bibfnamefont {K.}~\bibnamefont {Kostanjevec}},
  \bibinfo {author} {\bibfnamefont {J.~M.}\ \bibnamefont {Collinson}}, \bibinfo
  {author} {\bibfnamefont {R.}~\bibnamefont {Sknepnek}},\ and\ \bibinfo
  {author} {\bibfnamefont {E.}~\bibnamefont {Bertin}},\ }\bibfield  {title}
  {\bibinfo {title} {Dense active matter model of motion patterns in confluent
  cell monolayers},\ }\href@noop {} {\bibfield  {journal} {\bibinfo  {journal}
  {Nat. Commun.}\ }\textbf {\bibinfo {volume} {11}},\ \bibinfo {pages} {1405}
  (\bibinfo {year} {2020})}\BibitemShut {NoStop}%
\bibitem [{\citenamefont {Caprini}\ \emph {et~al.}(2020)\citenamefont
  {Caprini}, \citenamefont {Marconi}, \citenamefont {Maggi}, \citenamefont
  {Paoluzzi},\ and\ \citenamefont {Puglisi}}]{CapriniPRE20}%
  \BibitemOpen
  \bibfield  {author} {\bibinfo {author} {\bibfnamefont {L.}~\bibnamefont
  {Caprini}}, \bibinfo {author} {\bibfnamefont {U.~M.~B.}\ \bibnamefont
  {Marconi}}, \bibinfo {author} {\bibfnamefont {C.}~\bibnamefont {Maggi}},
  \bibinfo {author} {\bibfnamefont {M.}~\bibnamefont {Paoluzzi}},\ and\
  \bibinfo {author} {\bibfnamefont {A.}~\bibnamefont {Puglisi}},\ }\bibfield
  {title} {\bibinfo {title} {Hidden velocity ordering in dense suspensions of
  self-propelled disks},\ }\href
  {https://doi.org/10.1103/PhysRevResearch.2.023321} {\bibfield  {journal}
  {\bibinfo  {journal} {Phys. Rev. Research}\ }\textbf {\bibinfo {volume}
  {2}},\ \bibinfo {pages} {023321} (\bibinfo {year} {2020})}\BibitemShut
  {NoStop}%
\bibitem [{\citenamefont {Caprini}\ and\ \citenamefont {Marini
  Bettolo~Marconi}(2021)}]{CapriniSM21}%
  \BibitemOpen
  \bibfield  {author} {\bibinfo {author} {\bibfnamefont {L.}~\bibnamefont
  {Caprini}}\ and\ \bibinfo {author} {\bibfnamefont {U.}~\bibnamefont {Marini
  Bettolo~Marconi}},\ }\bibfield  {title} {\bibinfo {title} {Spatial velocity
  correlations in inertial systems of active brownian particles},\ }\href
  {https://doi.org/10.1039/D0SM02273J} {\bibfield  {journal} {\bibinfo
  {journal} {Soft Matter}\ }\textbf {\bibinfo {volume} {17}},\ \bibinfo {pages}
  {4109} (\bibinfo {year} {2021})}\BibitemShut {NoStop}%
\bibitem [{\citenamefont {Szamel}\ and\ \citenamefont
  {Flenner}(2021)}]{szamel2021longranged}%
  \BibitemOpen
  \bibfield  {author} {\bibinfo {author} {\bibfnamefont {G.}~\bibnamefont
  {Szamel}}\ and\ \bibinfo {author} {\bibfnamefont {E.}~\bibnamefont
  {Flenner}},\ }\bibfield  {title} {\bibinfo {title} {Long-ranged velocity
  correlations in dense systems of self-propelled particles},\ }\href
  {https://doi.org/10.1209/0295-5075/133/60002} {\bibfield  {journal} {\bibinfo
   {journal} {{EPL} (Europhysics Letters)}\ }\textbf {\bibinfo {volume}
  {133}},\ \bibinfo {pages} {60002} (\bibinfo {year} {2021})}\BibitemShut
  {NoStop}%
\end{thebibliography}%


\providecommand{\noopsort}[1]{}\providecommand{\singleletter}[1]{#1}%
\begin{thebibliography}{1}
\expandafter\ifx\csname natexlab\endcsname\relax\def\natexlab#1{#1}\fi
\expandafter\ifx\csname bibnamefont\endcsname\relax
  \def\bibnamefont#1{#1}\fi
\expandafter\ifx\csname bibfnamefont\endcsname\relax
  \def\bibfnamefont#1{#1}\fi
\expandafter\ifx\csname citenamefont\endcsname\relax
  \def\citenamefont#1{#1}\fi
\expandafter\ifx\csname url\endcsname\relax
  \def\url#1{\texttt{#1}}\fi
\expandafter\ifx\csname urlprefix\endcsname\relax\def\urlprefix{URL }\fi
\providecommand{\bibinfo}[2]{#2}
\providecommand{\eprint}[2][]{\url{#2}}

\bibitem[{\citenamefont{Abramowitz and Stegun}(1964)}]{abramowitz+stegun}
\bibinfo{author}{\bibfnamefont{M.}~\bibnamefont{Abramowitz}} \bibnamefont{and}
  \bibinfo{author}{\bibfnamefont{I.~A.} \bibnamefont{Stegun}},
  \emph{\bibinfo{title}{Handbook of Mathematical Functions with Formulas,
  Graphs, and Mathematical Tables}} (\bibinfo{publisher}{Dover},
  \bibinfo{address}{New York}, \bibinfo{year}{1964}).

\end{thebibliography}

\end{document}



\title{Supplementary Material: Kinetic Monte-Carlo Algorithms for Active-Matter systems}

\author{Juliane U. Klamser}
\email{juliane.klamser@espci.psl.eu}
\affiliation{Gulliver Lab, UMR CNRS 7083, PSL Research University, ESPCI Paris 10 rue Vauquelin, 75005 Paris, France}%

\author{Olivier Dauchot}%
\email{olivier.dauchot@espci.fr}
\affiliation{Gulliver Lab, UMR CNRS 7083, PSL Research University, ESPCI Paris 10 rue Vauquelin, 75005 Paris, France}%

\author{Julien Tailleur}
\email{julien.tailleur@u-paris.fr}
\affiliation{Universit\'e de Paris, Laboratoire Mati\`ere et Syst\`emes Complexes (MSC), UMR 7057 CNRS, F-75205 Paris, France}
\date{\today}


\maketitle

\tableofcontents

\section{Numerical implementation in 2d and persistence time}
\renewcommand{\thefigure}{S1}

\subsection{Numerical implementation of the reflecting boundary condition for the evolution of the self-propulsion speed}

\begin{figure}[h]
\centering
\subfloat{\label{fig1a}{\includegraphics[width=1\textwidth]{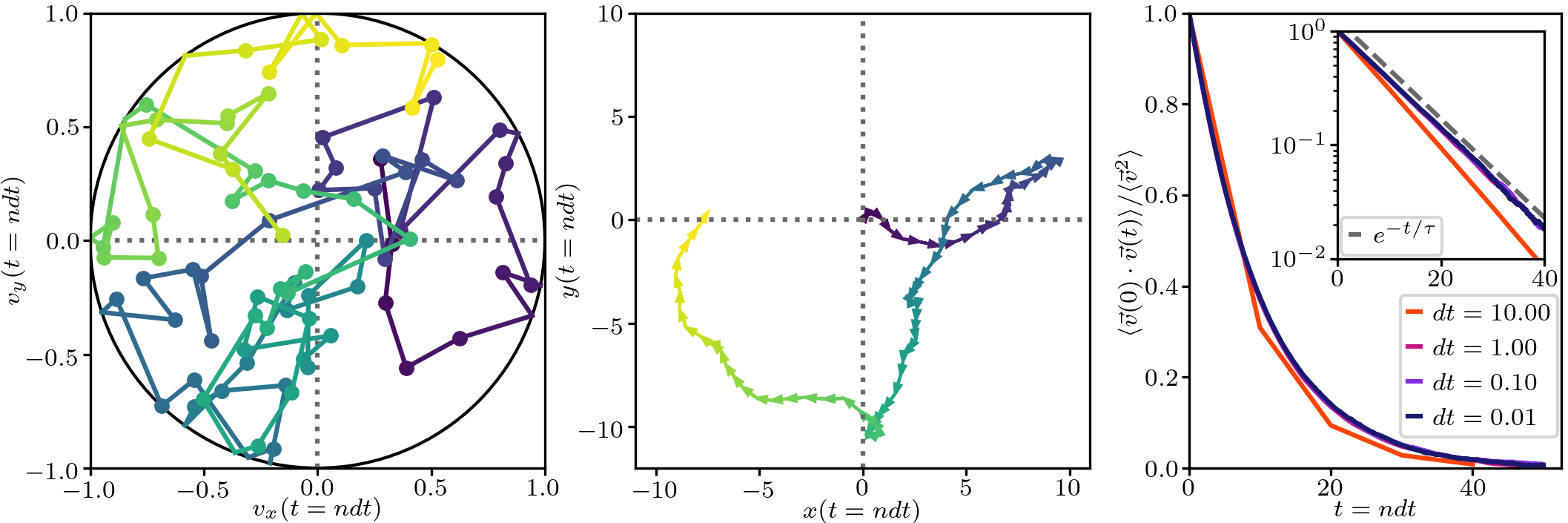}  }}%
\put(-510,0){(a) }
\put(-320,0){(b) }
\put(-150,0){(c) }
\caption{\textbf{(a)} Random walk of $\bv$ in the presence of circular
  reflecting boundaries at $|\bv| = 1$, $\tau = 10$ and $dt = 1$. The
  symbols correspond to the successive values of ${\bf v}(t_n)$. The
  color encodes progress in time from dark to light shades in (a) and
  (b). \textbf{(b)} The particle trajectory in position space
  corresponding to (a). \textbf{(c)} Normalized velocity
  auto-correlation for the decreasing values of $dt$ indicated in the
  legend. Remaining parameters as in (a). Overlapping curves show a
  fast convergence as $dt\to 0$. The semilogarithmic scale in the
  inset highlight the validity of the analyitical
  expression~\eqref{Eq:PerTime} for the persistence time. \label{F:velDyn} }
\end{figure}

The discrete-time implementation of the velocity dynamics is a random
walk with circular reflecting boundaries as illustrated in
Fig.~\ref{F:velDyn}a. In this random walk, the new velocity
$\bv(t_{n+1})$ is sampled from a Gaussian distribution
centered at $\bv(t_{n})$, of standard deviation $\delta v=\sqrt{2
  D_v dt}$.  The reflecting boundary conditions are implemented as
elastic reflections at $|\bv|=v_0$ that are applied when the
point sampled by the Gaussian is located outside the circle (see
Fig.~\ref{F:velDyn}a). 

For a free particle, the Metropolis filter in {Eq.~(1)} of the
main text is one, \textit{i.e.} every particle displacement is
accepted $\br(t_{n+1})= \br(t_{n})+dt\,\bv(t_{n})
$. Fig.~\ref{F:velDyn}b shows a discrete-time trajectory where
persistent motion is evident.

In the limit of $dt \to 0$, the random walk of $\bv$ converges to the
Langevin dynamics~(12) in the main text. The corresponding persistence
time is computed below, and given in Eq.~\eqref{Eq:PerTime}.
Figure~\ref{F:velDyn}c shows a rapid convergence of the discrete-time
algorithm to his asymptotic value for the persistence time.

\subsection{Computation of the persistence time}

In the continuous-time limit, the velocity dynamics is:
\begin{equation}
\dot\bv = \sqrt{2D_v} \,\boldsymbol \zeta \,,
\label{Eq:dotV2d}
\end{equation}
with reflecting boundary conditions at $|\bv|=v_0$ and $\boldsymbol
\zeta$ an uncorrelated unitary Gaussian white noise. To compute the
persistence time $\tau$, we consider the corresponding diffusion
equation in polar coordinates $(v,\theta)$,\begin{subequations}
\begin{equation}
\partial_t P_t( v, \theta) = D_v \Delta P_t( v, \theta )
\label{Eq:VelocityDiffusionEquation}
\end{equation}
with a zero current condition at $v = v_0$,
\begin{equation}
\partial_v P_t( v, \theta ) \mid_{v = v_0} = 0\,.
\label{Eq:Neumann}
\end{equation}
\label{Eq:Polar2DDiffusionEq}
\end{subequations}
The corresponding propagator is
\begin{equation}
P_t( v, \theta | v', \theta')  = \frac{1}{\pi v_0^2} + \sum_{m = 1}^\infty \sum_{n = 1}^\infty  e^{ -t/\tau_{mn} } J_m\left(\frac{s_{mn}}{v_0} v\right) \left[A_{mn} \cos(m \theta) + B_{mn} \sin(m \theta) \right]\,,
\label{Eq:P_t(v,theta)AB}
\end{equation}
where $\tau_{mn} = \frac{v_0^2}{s_{mn}^2 D_v}$, $J_m(x)$ is the
$m^{\text{th}}$ Bessel function of the first
  kind~\cite{abramowitz+stegun}, and $s_{mn}$ is the $n$-th zero of
  $\partial_xJ_m(x)$, which follows from the Neumann boundary
  conditions in Eq.~\eqref{Eq:Neumann}.  The constants $A_{mn}$ and
  $B_{mn}$ are functions of the initial velocity $(v', \theta')$.

We compute the persistence time from the exponential tail of the velocity autocorrelation in the stationary state
\begin{equation}
\begin{aligned}
\langle \vec{v}_2(t_2)\vec{v}_1(t_1)\rangle &= \int _{0 }^{2\pi }d\theta_1 d\theta_2\int _{0 }^{v_0 } dv_1 dv_2\, v_1 v_2  \cos(\theta_2 - \theta_1)P_{t_2-t_1}( v_2, \theta_2| v_1, \theta_1) P_{ss}(v_1, \theta_1)\,,
\end{aligned}
\label{Eq:AutoCorrVelocity}
\end{equation}
where $P_{ss}(v_1, \theta_1)= \tfrac{1}{\pi v_0^2}$ is the uniform stationary distribution. For large $t_2-t_1$, the leading-order contribution comes from the term $(m,n)\equiv (1,1)$ in Eq.~\eqref{Eq:P_t(v,theta)AB}. This gives $\langle \vec{v}_2(t_2)\vec{v}_1(t_1)\rangle \sim e^{-(t_2-t_1)/ \tau}$\,,
where the auto-correlation time
\begin{equation}
\tau = \tau_{1,1} = \frac{v_0^2}{c^2 D_v}\,,
\label{Eq:PerTime}
\end{equation}
with $c \equiv s_{1,1}\simeq1.841$.

\section{Simulation parameters}
{\bf Fig.~1a:} $\alpha = 1$, $\beta = 1$, $v_0 = 4$, $D_v=4.72$, leading to
$\tau \simeq 1$. $N=2000$ particles were simulated in a box of size
$L_x \times L_y$, where $L_y=35$ and $L_x=158.5$. Periodic boundary
conditions are used along the $y$ axis whereas hard walls are
implemented at $x=-69.5$ and $x=89$. The asymmetric potential is given
by $V(x,y) = 160 (x + 0.5)$ for $x \in [ -0.5, 0 ]$ and
$V(x,y) = 4 (20 - x)$ for $x \in [ 0, 20 ]$.

\medskip

{\bf Fig.~1b:} $\alpha = 1$, $\beta = 1$,  $v_0 = 1$, $D_v={1.475 \times 10^{-3}}$, leading to
$\tau \simeq 200$. $N=43\,904$ particles were simulated in a box of size
$L \times L$, where $L=468$. Periodic boundary
conditions are used along $x$ and $y$ axes.

\medskip

\textbf{Fig.~1c:} $\alpha = 1$, $\beta = 1$,  $v_0 = 1$, $D_v={2.95\times 10^{-2}}$, leading to
$\tau \simeq 10$. Periodic boundary conditions are used along the $y$
axis with $L_y=32$. The particles are confined along the $x$ axis at
$x=\pm 64$ by potentials $U_{\rm w}(x,y) = \tfrac{10}{8}(x\pm 64)^8$ for
$|x| > 64$.

\bigskip

\textbf{Fig.~2a:} $\alpha = 0.4$, $\beta=1$, $D = 1$, $v_0 = 4$, $D_v=4.72$ leading to $\tau \simeq 1$. $N=2000$
particles were simulated in a box of size $L_x\times L_y$, where
$L_y=35$ and $L_x=158.5$. Periodic boundary
conditions are used along the $y$ axis whereas hard walls are
implemented at $x=-69.5$ and $x=89$. The asymmetric obstacle is modelled by a potential
$V(x,y) = 18 (x + 0.5)$ for $x \in [ -0.5, 0 ]$ and
$V(x,y) = 0.45 (20 - x)$ for $x \in [ 0, 20 ]$.

\medskip

\textbf{Fig.~2b:} $\alpha = 0.6$, $\beta = 1$, $D = 0.05$,
$v_0 = 1$, $D_v={1.475 \times 10^{-3}}$ so that $\tau \simeq 200$. $N = 43\,904$
particles were simulated in a box of size $L \times L$, where
$L=270$. Periodic boundary conditions were used
along $x$ and $y$ axes.

\medskip

\textbf{Fig.~2c:} $\alpha = 0.6$, $\beta = 1$, $D = 0.05$,
$v_0=1$, $D_v={2.95 \times 10^{-2}}$ so that $ \tau \simeq 10$.  Periodic boundary
conditions are used along the $y$ axis with $L_y=32$. The particles
are confined along the $x$ axis at $x=\pm 64$ by potentials
$U_{\rm w}(x,y) = \tfrac{10}{8}(x\pm 64)^8$ for $|x| > 64$.

\bigskip

\textbf{Fig.~3a:} $\beta=1$, $D =1$, $v_0 = 1$, $D_v={8.1055\times 10^{-2}}$ so
that $\tau \simeq 5$, $dt = 10^{-4}$. The confining potential is given by
$U_{\rm w}(x,y) = \tfrac{\Omega}{\nu}(x\pm15)^\nu$ for $|x| > 15$ and the
different data sets correspond to
\begin{itemize}
\item $U_1: \nu = 8, \Omega = 10$,
\item $U_2: \nu = 6, \Omega = 10$,
\item $U_3: \nu = 4, \Omega = 10$,
\item $U_4: \nu = 2, \Omega = 100$,
\item $U_5: \nu = 2, \Omega = 10$.
\end{itemize}
\medskip

\textbf{Fig.~3b:} $\beta=1$, $\hat{D} =1/6$, $v_0 = 1$, $D_v={8.1055\times 10^{-2}}$ so that $\tau \simeq 5$, $dt = 10^{-4}$. The confining potential is given by
$U_{\rm w}(x,y) = \tfrac{\Omega}{\nu}(x\pm5)^\nu$ for $|x| > 5$, with $(\nu, \Omega)$ as in   Fig.~3a.

\section{AKMC algorithms and persistent time in 1d}
We describe below the one-dimensional version of the blended AKMC; the
purely active AKMC that we initially describe in the Letter
corresponds to setting $\alpha=1$ in the text below.

In 1d, an isolated particle of position $x$ and velocity $v$ in the
presence of an external potential $U(x)$ evolves as follows.  In one
MC step, the particle attempts a move $x_n \to x_n + \Delta x$. With
probability $\alpha$ the displacement is active, i.e.  $\Delta x =
dt\,v_n/\alpha$, and with probability $1-\alpha$ the particle attempts
a passive move, i.e. $\Delta x = \xi$, with $\xi$ uniformly sampled in
$[-\sqrt{6Ddt/(1-\alpha)}, \sqrt{6Ddt/(1-\alpha)}]$. In both cases,
the move is accepted with a probability given by the Metropolis filter
$f(x_n, \Delta x) = \min[1, \exp(-\beta \{U(x_n+\Delta x)-
  U(x_n)\})]$. A new velocity $v_{n+1}$ is then sampled. As in 2d, the
velocity undergoes a random walk with reflecting boundary conditions
at $|v| = v_0$. The new velocity $v_{n+1}$ is sampled form a Gaussian
distribution centred at $v_{n}$ and of standard deviation $\delta v =
\sqrt{2D_vdt}$. When the resulting value $\hat v$ is outside the
interval $[-v_0, v_0]$, $v_{n+1}$ is obtained using the elastic
reflection at $|v| = v_0$:
\begin{equation}
v_{n+1} = \begin{cases}
q- v_0\,,\text{ if } q < 2v_0\\
3v_0-q\,,\text{ if } q \geq 2v_0\\
\end{cases}\,,
\end{equation}
where $q = (\hat v+v_0)\,\text{mod}\,(4v_0)$. Here we use the definition of the modulo as $a \,\text{mod}\, b = a - b \floor{\frac{a}{b}}$ with $\floor{a}$ denoting the floor function. Therefore, $q$ satisfies $0\leq q \leq 4v_0$ and thus $|v_{n+1}| < v_0$. 

In 1d, the velocity dynamics describing the blended AKMC in the
continuous-time limit is given by $\dot v = \sqrt{2D_v}\zeta$, with
additional reflecting boundaries at $|v| = v_0$, and $\zeta$ is again
a unitary Gaussian white noise.

The derivation of the persistence time in 1d is analogous to the 2d calculation, leading to the corresponding propagator
\begin{equation}
P_t( v | v' )  = \frac{1}{2 v_0} + \frac{1}{v_0}  \left(\sum_{n = 2,4,6}^\infty   A_{n} \cos\left(\frac{n \pi v}{2v_0} \right) e^{ -t / \tau_n } + \sum_{n = 1,3,5}^\infty   B_{n} \sin\left(\frac{n \pi v}{2v_0} \right) e^{ -t / \tau_n }\right) \,,
\label{Eq:P_t(v,theta)AB_1d}
\end{equation}
with
\begin{equation}
\tau_n = \frac{4v_0^2}{n^2 \pi^2 D_v}\,,
\end{equation}
and the constants $A_n$ and $B_n$ are functions of the initial $v'$.
Following a similar argument as in 2d, the auto-correlation time $\tau
\equiv \tau_1$ as given in the Letter.

\bibliography{biblio}